\documentclass{aa}
\usepackage{txfonts}
\usepackage{graphicx}

\usepackage{amsfonts}

\def\a{\alpha}

\def\d{\delta}

\def\r{\rho}

\def\g{\gamma}

\def\th{\theta}

\def\s{\sigma}
\def\ra{\rightarrow}
\def\Ra{\Rightarrow}
\def\Lra{\Leftrightarrow}

\def\Mfunction#1{\mathop{\rm #1}\nolimits}

\def\bm#1{\mbox{\boldmath{$#1$}}}

\begin{document}

\title{Geometry and scaling of cosmic voids}

\author{Jos\'e Gaite}
\institute{Instituto de Microgravedad IDR, 
ETS Ingenieros Aeron\'auticos, 
Universidad Polit\'ecnica de Madrid,
E-28040 Madrid, Spain; jose.gaite@upm.es
}

\date{\today}

\abstract{Cosmic voids are observed in the distribution of galaxies and, to
  some extent, in the dark matter distribution. If these distributions have
  fractal geometry, it must be reflected in the geometry of voids; in
  particular, we expect scaling sizes of voids.  However, this scaling is not
  well demonstrated in galaxy surveys yet.  }
{Our objective is to understand the geometry of cosmic voids and its relation
to the geometry of the galaxy and dark-matter distributions.  We examine the
consequences of a fractal structure of matter and, in particular, the
hypothesis of scaling of voids.  We intend to distinguish monofractal voids
from multifractal voids, regarding their scaling properties.  We plan to
analyse voids in the distributions of mass concentrations (halos) in a
multifractal and their relation to galaxy voids.  }
{We begin with a statistical analysis of point distributions based on the void
probability function and correlation functions. An analytical treatment is
possible if we assume that voids are spherical.  Therefore, we devise a simple
spherical void finder.  For continuous mass distributions, we employ the
methods of fractal geometry.  These methods provide analytical predictions,
which we confirm with numerical simulations.  Smoothed mass distributions are
suitable for the method of excursion sets.  }
{Voids are very nonlinear and non-perturbative structures.  If the matter
distribution has fractal geometry, voids reflect it, but not always directly:
scaling sizes of voids imply fractal geometry, but fractal voids may have a
complicated geometry and may not have scaling sizes.  Proper multifractal
voids are of this type.  A natural multifractal biasing model implies that the
voids in the galaxy distribution inherit the same complicated geometry.  }
{Current galaxy surveys as well as cosmological $N$-body simulations indicate
that cosmic voids are proper multifractal voids.  This implies the presence in
the voids of galaxies or, at least, small dark matter halos.}  

\maketitle

\section{Introduction}

The large scale structure of matter is formed by clusters, filaments, sheets,
and voids.  The r\^ole of voids as basic ingredients of the large scale
structure is now well established, but the definition of what constitutes a
void is still imprecise.  Originally, voids were described as large regions
devoid of galaxies, but the current view is more complex (Peebles, 2001).
Consequently, the theory has developed progressively to consider more
sophisticated models of voids.  Nowadays, there is a good amount of statistics
of galaxy voids, and there is also information about voids as a part of the
more general information about cosmic structure that is obtained in $N$-body
simulations of cold dark matter (CDM).  It seems that it should be possible to
determine the main features of voids by combining observational data with
theoretical models.

Here, we are interested in the features of voids related to the scale
invariance and fractal geometry of the cosmic structure.  Scale invariance is
a general symmetry, which can manifest itself in physical systems in various
ways.  In cosmology, it gave rise to hierarchical models of the Universe,
related to fractal geometry (Mandelbrot, 1983). Some evidence for them was
provided by the analysis of galaxy catalogues, which found that the two-point
correlation function is a power law on scales of several Mpc, and found, with
less confidence, that higher-order correlation functions are also power laws
(Peebles, 1980).  Self-similar fractal models of the galaxy distribution have
been well studied (Pietronero, 1987; Sylos Labini, Montuori \& Pietronero,
1998); also, multifractal models (Jones et al, 1988; Balian \& Schaeffer,
1988; Mart{\'\i}nez, Jones, Dom{\'\i}nguez-Tenreiro \& van de Weygaert, 1990).
Recent reviews of various ideas and models in cosmology based on scaling laws
are given by Jones et al (2004) and by Gabrielli et al (2005).

Mandelbrot (1983) introduced the notion of fractal holes (``tremas'') and
considered its application to the galaxy distribution.  In fact, he was
concerned with the absence of large voids in this distribution and introduced
the concept of lacunarity in this regard: two fractals with equal dimension
can differ in their lacunarity, in such a way that the less lacunar one
appears as if it had larger dimension (roughly speaking, as if it were less
fractal).  Besides, Mandelbrot (1983) characterized some fractals as having a
power-law distribution of void sizes.

Regarding astronomical measures of voids, self-similarity of voids was already
considered by Einasto et al (1989) as a probe of scale invariance in the large
scale structure.  Following Mandelbrot (1983), we proposed that a
manifestation of the self-similarity of voids, as it appears in fractal
distributions, is that the rank-ordering of their sizes fulfills Zipf's power
law (Gaite \& Manrubia, 2002). In particular, the transition from this power
law to a different dependence at small ranks, namely, the fact that the
largest voids have almost constant size, marks the transition to homogeneity
on very large scales (Gaite, 2005-B).  Thus far, the scaling of galaxy void
sizes remains moot. Our early analysis of void catalogues did not show
any evidence of a Zipf law (Gaite \& Manrubia, 2002), but analyses of recent
surveys are more favourable (Tikhonov \& Karachentsev, 2006; Tikhonov, 2006;
Tikhonov, 2007). However, scaling over a convincingly large range has not been
demonstrated yet.

Otto et al.\ (1986) regarded the significance of cosmic voids, namely, they
regarded whether voids were really elements of the large scale structure or
just the necessary fluctuations about homogeneity.  They developed statistical
methods to answer this question.  These methods were improved and generalized
by Betancort-Rijo (1990).  Our statistical methods are based on theirs, but
our constructions are addressed to the study of the nonlinear regime and, in
particular, the analysis of its scale invariant features.

The detection of voids in galaxy samples is carried out with computer
algorithms called void-finders. These algorithms have evolved in accord with
the observational and theoretical ideas about voids, in particular, in accord
with the ideas about their emptiness.  Initially, voids were defined as empty
spheres of maximal radii (Einasto et al, 1989). Later, more general shapes
were allowed, and voids were also allowed to contain some galaxies, as in,
e.g., the popular void-finder devised by El-Ad \& Piran (1997), which
separates ``field galaxies'' from ``wall galaxies.''  We have defined an
algorithm for finding voids of arbitrary shape based on discrete-geometry
constructions (Gaite, 2005-B).  Here, we return to the old spherical voids,
which are more adequate for the application of analytical methods. In
particular, they are adequate for designing a parameter-free void-finder.
Spherical voids are also favoured by theoretical arguments (Icke, 1984).

There is no substantial observational knowledge of the geometry of voids in
the dark matter distribution.  Notwithstanding, voids have been studied in
cosmological $N$-body simulations of CDM models. Gottl\"ober et al (2003) have
re-simulated voids with higher resolution and found structures inside them, in
a self-similar pattern.  Recently, Sheth \& van de Weygaert (2004) and
Shandarin, Sheth \& Sahni (2004) have defined voids as {\em under-dense
connected} regions, such that they are complementary to clusters. With this
definition, voids contain matter and have very complex shapes.

The hypothesis of scale invariance in the nonlinear regime leads to a
self-similar multifractal model of the dark matter distribution, which is
supported by the analysis of cosmological simulations (Gaite, 2005-A, 2007).
It is natural to define voids in a multifractal as locations of mass
depletions (Gaite, 2007).  We study here the geometry of these voids, which is
very complicated, even more complicated than the geometry of the voids in the
models of Sheth \& van de Weygaert (2004) or Shandarin, Sheth \& Sahni (2004).
Assuming that the dark-matter distribution is multifractal, it is desirable to
relate multifractal voids, which have complex shapes and may contain matter,
to simpler definitions applicable to galaxy voids (which are the ones that we
can directly observe). Two questions are important in this regard: (i) the
role of the small number density the distribution of galaxies in comparison
with the distribution of dark matter, which can be considered continuous; (ii)
the statistical relation of the distribution of galaxies to the distribution
of dark matter, which can be direct or involve galaxy biasing.

Regarding the number density of galaxies, we study how the nature of voids in
a sample of a continuous distribution changes with the density of the
sampling, especially, in the nonlinear regime.  Regarding galaxy biasing, we
only consider a simple model, which relies on the ``peak theory'' of Gaussian
fields (Kaiser, 1984; Bardeen et al, 1986). However, we introduce the
non-Gaussian character of the density field in the nonlinear regime by
substituting Gaussian peaks by nonlinear mass concentrations (halos), as
proposed before (Gaite, 2005-A, 2007).  Therefore, we can apply our
results about scaling in the distribution of halos of given mass in a
multifractal cosmic distribution.

We begin with a brief review of the notion of a void in a discrete mass
distribution.  We call this type of voids Poissonian voids.  Indeed, a
complete analytical study of this type of voids can be carried out for a
Poisson distribution (Sect.\ \ref{Poisson}). This study is useful for setting
up the analytical framework and for establishing the significance of voids.
The generalization to correlated point distributions yields perturbative
expressions in terms of correlation functions (Sect.\ \ref{correl}).  The
study of Poissonian voids is simplified by the definition of a void as an {\em
empty} spherical region enclosed by a set of four non-coplanar points. Relying
on this definition and imposing the condition that the voids do not overlap,
we devise a simple and efficient void-finder (Sect.\ \ref{void-f}).  Then, we
consider the scaling of voids in monofractal distributions (Sect.\
\ref{scaling}).  We review the definition and properties of cut-out sets
as cosmic foam models (Gaite, 2006) and generalize that notion.  With this
background, we undertake the general case of multifractal voids, which leads
us to differentiate two types of voids, in connection with their geometry
(Sect.\ \ref{MF}).  Then, we study galaxy voids and galaxy biasing (Sect.\
\ref{bias}).  Finally, we discuss our results (Sect.\ \ref{discuss}).

\section{Poissonian analysis of galaxy voids}
\label{Poisson}

Voids arose as large regions in the distribution of galaxies containing no
galaxies or many fewer than the mean expected number of galaxies. Of course,
there are fluctuations even in a homogeneous distribution of
galaxies. Therefore, it is necessary to quantify the fluctuations.

To be precise, a homogeneous Poisson field (or process) is defined as a random
sample of a uniform distribution, such that the probability of having a point
in a given region is proportional to its volume. If the density (or intensity)
of the field is $n$, then the probability of having $k$ points in a region of
volume $V$ is given by the familiar Poisson distribution with parameter $N =
nV$:
$${P}_k[V] = \frac{N^k}{k!}\,e^{-N}.$$ 
Politzer \& Preskill (1986) studied the statistics of clusters and voids in a
Poisson field. This study was used by Otto et al.\ (1986) to determine the
significance of the then observed galaxy voids.  Their conclusion is that
those voids were consistent with a homogeneous distribution of galaxies or, at
any rate, with a homogeneous distribution of rich clusters of galaxies.

Central to the arguments of Otto et al.\ (1986) is the calculation by Politzer
\& Preskill (1986) of the probability per unit volume of having a void of
given (and simple) shape in a random sample of points with a uniform
distribution. The probability that a {\em given} region contains no points or
contains fewer than the expected number of points is provided by the Poisson
distribution. However, the calculation of the probability of having a void of
given size and shape at any place is a more difficult problem.  Its solution
can be obtained with a careful analysis of the Poisson field. Their analysis
is involved and their formula for the probability of a void has a questionable
normalization.  Since this formula is important, we re-derive it (with its
normalization) using a different method that is more straightforward. Then, we
re-analyse the significance of voids, regarding the voids in some recent
galaxy surveys and their statistics.

\subsection{Probability of a spherical void in a homogeneous distribution}
\label{sph-void}

The simplest shape of a void is certainly the spherical shape.  Politzer \&
Preskill (1986) obtained the following formula for the probability per unit
volume of a spherical void:
$$
\mathbb{P}_k[V] = \left(\frac{3\pi^2}{32}\right)
\frac{(nV)^3}{V}\,{P}_k[V]\,,
$$
where $k$ is the number of points in the void, assumed to be much smaller than
the expected number, namely, $k \ll nV$.

The probability density $\mathbb{P}_k[V]$ should be normalized, that is to
say, its integral over $V$ from zero to infinity should be one. However, the
constant ${3\pi^2}/{32}$ does not normalize it, as is easily checked.  Since
Politzer \& Preskill's formula for $\mathbb{P}_k[V]$ is only valid for $V \gg
k/n$, it cannot be normalized. Nevertheless, that condition is certainly
fulfilled for all $V > 0$ if $k=0$, which suggests that the probability of an
empty void $\mathbb{P}_0[V]$ should be valid for all $V$.

If we assume that Politzer \& Preskill's formula holds for all $V$, then the
constant ${3\pi^2}/{32}$ must be replaced with
$1/\left((k+2)(k+1)\right)$. Note that, then, only $\mathbb{P}_0[V]$ (totally
empty voids) refers to voids over the entire range of $V$.  Using a method
simpler than the one of Politzer \& Preskill, we can prove that, in fact,
\begin{equation}
\mathbb{P}_k[V] = \left(\frac{1}{(k+2)(k+1)}\right)
\frac{(nV)^3}{V}\,{P}_k[V]\,,
\label{cPP}
\end{equation}
is correct and holds over the entire range of $V$.

\subsubsection{Calculation of the probability $\mathbb{P}_k[V]$}
\label{Pk}

The calculation of $\mathbb{P}_k[V]$ for spherical voids can be formulated as
follows. A spherical void is defined by four non-coplanar points on its
boundary, because a void can always be enlarged so as to touch four points. In
addition a $k$-void is defined to have $k$ points inside. Since the points are
uncorrelated, the probability distribution for each point is constant and
independent of the other points. Therefore, we can calculate $\mathbb{P}_k[V]$
as the product of the probability of four points and the conditional
probability of having $k$ points inside the sphere of volume $V$ defined by
them, ${P}_k[V]$.

The probability of each of the four points is the product of its volume
element and the density $n$. Thus, the only problem is to express the
four-point volume element in terms of a set of variables that includes the
volume of the sphere defined by the four points. If we denote the positions of
the points by $\{x_i\}_{i=1}^4$ and the position of the center of the sphere
(their circumcenter) by $x_{\rm c}$, then
\begin{eqnarray*}
d^3x_1 \,d^3x_2\,d^3x_3\,d^3x_4 =\\ d^3x_{\rm c}\,V^2\,dV\,
f({\bm \th}_1,{\bm \th}_2,{\bm \th}_3,{\bm \th}_4)\,
d^2{\bm \th}_1\,d^2{\bm \th}_2\,d^2{\bm \th}_3\,d^2{\bm \th}_4,
\end{eqnarray*}
where $V$ is the sphere's volume, $\{{\bm \th}_i\}_{i=1}^4$ are the four sets
of two angular coordinates over the sphere, and $f$ is a function of these
angular coordinates.  This expression follows from translation invariance and
dilation covariance only. The function $f$ can be calculated, but we do not
need it.  We obtain formula~(\ref{cPP}), without normalization, by factoring
out both the integral over the angles and $n\,d^3x_{\rm c}$.  A
straightforward integration over $V$ gives the correct normalization constant
$1/\left((k+2)(k+1)\right)$.

Let us remark that formula~(\ref{cPP}) involves {\em no approximation} and is
valid for all $V$. However, it refers to voids only if $k \ll nV$.  In the
case of empty voids ($k=0$), then ${P}_0[V]$ approaches unity in the limit $V
\ra 0$, but $\mathbb{P}_0[V]$ is small due to the boundary factor
(corresponding to the four points defining the void). In other words, the
probability that a randomly chosen small ball be empty is large, but the
probability that it constitute a void is small. In fact, the most probable
void size is $V=2/n$.  We can also calculate other statistical quantities, for
example, the average void size, ${\overline V}=3/n$.

\subsection{The significance of large voids}
\label{signif}

Otto et al.\ (1986) used Politzer \& Preskill's formula to determine the
significance of some large voids that had been found at the time. In
particular, they chose a totally empty void in the distribution of Abell
clusters (rich clusters), in which the expected number of these clusters was
10. Then, Politzer \& Preskill's formula, with $k=0$ and $nV=10$, yields
$\mathbb{P}_0[V] = 0.042/V$. Multiplying it by the total volume, which
corresponds to a sample of 70 clusters, they obtain that the expected number
of voids of that size is $0.3$.  However, if we replace the constant
${3\pi^2}/{32} = 0.93$ with its correct value, $1/2$, then the expected number
of voids is halved, so that the void in question becomes less probable.

On the other hand, the actual void was ellipsoidal. Thus, we should consider
the largest spherical void inscribed in it, which is quite smaller and,
therefore, more probable. Alternately, we could use the formula given by Otto
et al.\ (1986) for ellipsoidal voids.\footnote{In this case, Otto et al.\ do
not give the normalization constant, which depends on the maximum accepted
eccentricity. This constant can be obtained with our method.}  At any rate,
their conclusion was that such a void was not improbable even in a random
distribution of Abell clusters. This negative conclusion needs revision,
anyway, regarding recent data.  In particular, we shall briefly regard below
data from the 2dF and SDSS galaxy surveys that show that large voids are, in
fact, very significant. Of course, this is the commonly accepted conclusion
nowadays.

Let us explain in more detail the use of Eq.~(\ref{cPP}) to determine the
departure of a point distribution from randomness. Since we are interested in
large voids, we can use the cumulative probability of having a void of size
equal or larger than $V$:
$$
\mathbb{P}_>[V] = 
\int_V^\infty\mathbb{P}_0[v]\,dv \approx \frac{(nV)^2}{2}\,e^{-nV},
$$ 
for large voids ($nV \gg 1$). Setting $nV = 10$ (as Otto et al.),
$\mathbb{P}_> = 0.0023$, which is a fairly small number. However, the expected
number of voids of size larger than $V$ is naturally proportional to the size
of the sample. Thus, it is a number relatively close to one for a moderate
sample of 70 points and, in fact, it becomes larger than one for a large
enough sample.  Nevertheless, in a large sample, the natural question is if
there are other large voids and their respective sizes.

In other words, for a complete study of the significance of voids in a sample,
one must determine the sizes of many voids, beginning with the largest one.
This is the standard void-finding procedure. Suitable void-finding algorithms
are employed in this task. If the void-finder looks for voids by just fitting
the largest sphere, then this spherical void should be the most
significant. Smaller voids found in the sample are usually constrained not to
overlap any preceding void (or not to overlap more than a given fraction of
any preceding void).  We will analyse a void-finder of this type in Sect.\
\ref{void-f}.  Once the full set of voids in a sample is available, one can
study the distribution of their sizes. In recent surveys, we usually find that
the largest voids are very significant, but we see below that the overall
distribution of void sizes can be even more significant.

\subsubsection{Voids in recent surveys}
\label{Tikho}

Tikhonov (2006, 2007) has studied and rank-ordered the voids in samples of the
2dF and SDSS surveys.  In particular, he has chosen from the 2dF survey a
volume limited sample (VLS) with 7219 galaxies, such that the volume per
galaxy is 513 Mpc$^3$ $h^{-3}$. The largest void in it corresponds to a sphere
with radius 21.3 Mpc $h^{-1}$ and volume $4.05~10^4$ Mpc$^3$ $h^{-3}$. Thus,
if the sample belonged to a uniform distribution, the expected number of
sample galaxies in that sphere would be 78.9.  Then, the expected number of
voids of that size would be $(1/2)\,7219\times 78.9^2\, e^{-78.9} =
1.21~10^{-27}$, that is, absolutely negligible.

Furthermore, Tikhonov (2006) finds a number of voids with corresponding
spheres that have slightly smaller radii. All these voids are undeniably
significant. Indeed, Tikhonov (2006) uses a numerical comparison with a random
sample to determine that the largest 110 voids, with sphere radius larger than
9 Mpc $h^{-1}$, are significant.  His criterion is to measure significance by
$1- N_{\rm random}(r)/N_{\rm survey}(r)$, where $N_{\rm survey}(r)$ and
$N_{\rm random}(r)$ are the number of voids with radii larger than $r$,
respectively, in the sample and in a random point distribution with the same
boundaries and mean density.  That quantity must be non-vanishing for
significant voids.  The mean number of galaxies in a sphere with radius of 9
Mpc $h^{-1}$ is six, and the expected number of larger void spheres in the
sample is $N_{\rm random}(6) = 460$.%
\footnote{ The smaller number reported by Tikhonov (2006), namely, 110, is due
to the non-overlap condition and to the fact that his voids are not spherical
and, therefore, they are larger.}  However, note that Tikhonov's criterion is
less stringent than Otto et al's, which requires $N_{\rm random}(r) \ll N_{\rm
survey}(r)$. In other words, Otto et al's criterion neglects voids that are
just somewhat larger than the voids in a random point distribution, because
the probability of such voids in a random point distribution is not
sufficiently small.

The very recent analysis of SDSS voids (Tikhonov, 2007) yields similar
results, regarding the size of the largest voids and, indeed, the overall
distribution.  We reanalyse in Sect.\ \ref{bias} Tikhonov's VLS from the 2dF
survey, regarding the statistical properties of the distribution of its voids,
in particular, its scaling properties.

\section{Voids in correlated point distributions}
\label{correl}

It is possible to extend the preceding methods to correlated point fields.%
\footnote{The precise formulation of this generalization actually involves
{\em two} stochastic processes: the point process and the random process that
produces the continuous distribution. Thus, it is called doubly stochastic
process or Cox process.}  Given that the distribution of galaxies is very
inhomogeneous on small scales, this extension is necessary.  Otto et al.\
(1986) already considered the modification of their results when the points
are correlated rather than totally random, and studied it in terms of the
cluster expansion.  Betancort-Rijo (1990) employed a different method, which
is more adequate when the correlations are strong.

We consider again the probability ${P}_k[V]$ of having $k$ points in a region
of volume $V$.  For the Poisson distribution, its maximum is at $k \simeq nV$
(which is also the mean). Naturally, the presence of density fluctuations
makes ${P}_k[V]$ larger when $k$ differs from $nV$, in particular, when $k \ll
nV$.  Actually, the basic quantity that we need is ${P}_0[V]$ (the void
probability function), because ${P}_k[V]$ can be obtained from it (White,
1979) and, besides, we are going to restrict ourselves to empty spherical
voids.

\subsection{The void probability
function and the probability of spherical voids}

Otto et al.\ (1986) relied on the expression of the void probability function
(VPF) in terms of correlations.  This function is analogous to the
grand-canonical partition function of a statistical particle system in which
their velocities have been integrated over. Therefore, the VPF admits an
expansion in terms of the correlations that is analogous to the cluster
expansion in statistical mechanics (see White 1979, for the introduction of
the VPF in cosmology, Balian \& Schaeffer 1989, for a more detailed study, and
Mekjian 2007, for the connection with statistical mechanics).

In analogy with the cluster expansion, if the correlation functions decay
rapidly on large distances, the void probability function can be expressed as
\begin{equation}
{P}_0[V] = \exp\left[\sum_{k=1}^{\infty} \frac{(-nV)^k}{k!}\, 
{\bar\xi}_k \right],
\label{vpf}
\end{equation}
where ${\bar\xi}_1 = 1$, and, for $k \geq 2$, 
$${\bar\xi}_k = \frac{1}{V^k} \int_V d^3x_1 \cdots d^3x_k\, \xi_k(x_1, \ldots,
x_k).$$ 
Note that Eq.~(\ref{vpf}) reduces to the Poisson form in the uncorrelated
case, namely, when ${\bar\xi}_k = 0$ for $k \geq 2$. 

The cumulants ${\bar\xi}_k, \;k \geq 2,$ vanish in the limit $V \ra
\infty$. Furthermore, in this limit, the larger $k$ is, the more rapidly they
vanish. Thus, the first approximation is to consider that only ${\bar\xi}_2$
is non-vanishing, that is to say, to consider a random sample of a Gaussian
field. The Gaussian approximation is reasonable, for example, when $\s =
{{\bar\xi}_2}^{1/2} < 0.32$, because the probability that the density be
negative is then smaller than 0.1\%. However, in the weakly-correlated regime
with $\s = 0.32$, the correction factor given by Eq.~(\ref{vpf}) is large even
for small $nV$. For example, when $nV = 4$, we have $\exp[(nV\s)^2 /2] = 2.3$,
and, when $nV = 10$ (as in the example in Sect.\ \ref{signif}), $\exp[(nV\s)^2
/2] = 167$.  Naturally, the presence of density fluctuations results in an
increase of the void probability function, an increase that can be large.

Let us mention that Otto et al.\ (1986), using the analogy with the cluster
expansion, interpret Eq.~(\ref{vpf}) as a renormalization of the number
density, such that
$$n_c =  n \left(1 +  \sum_{k=1}^{\infty} \frac{(-nV)^k}{(k+1)!}\, 
{\bar\xi}_{k+1} \right)$$
is the density of clusters.  Therefore, the statistics of voids on the scales
of transition to homogeneity in a correlated point distribution is similar to
the statistics of voids in an uncorrelated distribution with a lower density.
Within this interpretation, and in the Gaussian approximation with $\s = 0.32$
and $nV = 10$, we have that $n_c = n (1 - nV\s^2/2) = 0.49 n$. But $n_c$ turns
negative when $nV > 19$ and becomes meaningless.  Given that $n_c <0 \Ra
{P}_0[V] > 1$, we should not trust the Gaussian approximation if $\s$ {\em
and} $nV$ are not sufficiently small for having positive $n_c$, whether we
interpret it as a renormalized number density or not.

In a particular case of strong correlations, namely, in fractal distributions,
the transition to homogeneity in the statistics of voids has been studied by
Gaite (2005-B). The argument proposed there is that a fractal consists of a
hierarchy of clusters of clusters and, therefore, it is invariant under coarse
graining up to the homogeneity scale. Over this scale, the largest
coarse-grained particles (clusters) become uncorrelated and, correspondingly,
the largest voids are like voids in a Poisson distribution of
clusters. Simulations of fractals confirm this view (Gaite, 2005-B).

In the weakly nonlinear regime, an improvement over the Gaussian approximation
is the lognormal model (Coles \& Jones, 1991).  Its void probability function
is given by the probability of having a void volume $V$ in a density $\r$
integrated over the density; namely,
\begin{equation}
{P}_0[V] = \frac{1}{\s\sqrt{2\pi}}\int\limits_0^\infty 
\exp\left[-\frac{(\ln\r-\mu)^2}{2\s^2}-nV\r\right]
\frac{d\r}{\r},
\label{vpf-logn}
\end{equation}
where $\mu=-\s^2/2$ if we assume that $\langle\r\rangle = 1$.  This function
tends to the Gaussian VPF when $\s \ra 0$ (when the correlations become
small).  With $\s = 0.32$ and $nV = 10$, it yields ${P}_0[V] = 7.61\,10^{-4}$,
which is 17 times larger than the Poisson value (Sect.\ \ref{signif}).  This
factor is to be compared with the factor given by the Gaussian approximation,
namely, 167, which is clearly an overshot. Thus, this comparison questions the
validity of the Gaussian approximation well before $n_c$ becomes negative.

Once we have studied the VPF, we can generalize the calculation of
$\mathbb{P}_0[V]$ in Sect.\ \ref{Pk}. Let us begin with the expression for the
probability of having an empty sphere of volume $V$ with four non-coplanar
points on its boundary, namely, with the four-point function
\begin{eqnarray}
dP_{1234} = n^4
d^3x_1 \cdots d^3x_4 
\left[ {\bar\xi}(x_1) {\bar\xi}(x_2) {\bar\xi}(x_3) {\bar\xi}(x_4) + 
\right.\nonumber\\ \left.
{\bar\xi}(x_1,x_2) {\bar\xi}(x_3) {\bar\xi}(x_4) + \cdots +
{\bar\xi}(x_1) {\bar\xi}(x_2) {\bar\xi}(x_3,x_4) + 
\right.\nonumber\\ 
{\bar\xi}(x_1,x_2) {\bar\xi}(x_3,x_4) +
\cdots +
{\bar\xi}(x_1,x_4) {\bar\xi}(x_2,x_3) +
\nonumber\\
{\bar\xi}(x_1, x_2 ,x_3) {\bar\xi}(x_4) +
\cdots + {\bar\xi}(x_1) {\bar\xi}(x_2,x_3,x_4) + 
\nonumber\\
\left. 
{\bar\xi}(x_1, \ldots ,x_4) \right] {P}_0[V],
\label{Pclus}
\end{eqnarray}
where
\begin{eqnarray*}
{\bar\xi}(x) &=& 1 + 
\sum_{k=1}^{\infty} \frac{(-n)^k}{k!}
\int_V d^3y_1 \cdots d^3y_k\, 
\xi_{k+1}(x,,y_1, \ldots, y_k),\\
{\bar\xi}(x_1,x_2) &=& {\xi}_2(x_1,x_2) +\\
&&\sum_{k=1}^{\infty} \frac{(-n)^k}{k!}
\int_V d^3y_1 \cdots d^3y_k\, 
\xi_{k+2}(x_1,x_2,y_1, \ldots, y_k),
\end{eqnarray*}
and the other ${\bar\xi}(\cdot)$ are defined analogously (White, 1979; Balian
\& Schaeffer, 1989). Actually, ${\bar\xi}(x)$ is independent of $x$, due to
translation invariance, and $n\,{\bar\xi}(x)$ can be interpreted as a density
corrected by the presence of the neighbouring void.\footnote{This density is
related to the cluster density by $n{\bar\xi}(x) = n\frac{dn_c}{dn}.$} This
property can be used to rewrite Eq.\ (\ref{Pclus}) in a simpler fashion.

Following the procedure in Sect.\ \ref{Pk}, we can derive from Eq.\
(\ref{Pclus}) the expression for $\mathbb{P}_0[V]$.  Unfortunately, it is a
complex expression that depends in a complicated way on $V$ and the angular
coordinates, unlike in the Poisson case.  However, in the Gaussian
approximation ($V \ra \infty$), Eq.\ (\ref{Pclus}) simplifies and we can then
specify the corrections to the Poisson formula for $\mathbb{P}_0[V]$. Given
that ${\xi}_2$ is positive in the sphere, the correction to the density is
negative. Thus, the sign of the total correction to the Poisson formula
depends on the balance between this negative correction, the positive
corrections provided by ${\xi}_2(x_i,x_j), \; 1 \leq i,j \leq 4, \; i \neq j$,
and, in addition, the already considered correction to ${P}_0[V]$ (which is
positive).  We notice that this result for the Gaussian $\mathbb{P}_0[V]$ is
different from and more complicated than the formula given by Betancort-Rijo
(1990).

\subsection{The cumulant expansion of the VPF and the nonlinear regime}

We have seen that the Gaussian approximation fails when $N=nV$ grows. In
general, we may ask the radius of convergence of the power expansion in Eq.\
(\ref{vpf}). It turns out that this expansion is not necessarily convergent,
that is to say, its radius of convergence can be zero.  For example, the
expansion corresponding to the lognormal model is indeed not convergent but
only asymptotic as $N \ra 0$. In other words, it is only useful for
computations as long as there is a sufficient number of decreasing terms.  In
particular, for given ${\bar\xi}_2$, the (asymptotic) convergence of that
series is limited to $N < {\bar\xi}_2^{-1}$, namely, to the value of $N$ that
makes the magnitude of the second term of the series equal to the magnitude of
the first term (for larger $N$, the terms increase in absolute value).

We can interpret the condition $N < {\bar\xi}_2^{-1}$ as follows. Let us
consider the number variance in the volume $V$
$$
\frac{\langle \d N^2 \rangle}{N^2} = \frac{1}{N} + {\bar\xi}_2.
$$
The condition $N < {\bar\xi}_2^{-1}$ tell us that the Poisson fluctuations
dominate over the correlations. If these are small, as in a Gaussian
distribution, the Poisson fluctuations dominate, unless $N$ is relatively
large (when $N=10$, the corrections are already important in the Gaussian
example above, with $\bar\xi_2 = 0.32^2 \simeq 0.1$).  In contrast, in a
strongly correlated distribution $\bar\xi_2 \gg 1$, the Poisson fluctuations
are only important in very small volumes. 

We can illustrate the behaviour of the VPF cumulant expansion for the
lognormal model in the nonlinear regime with an example. Let $\s = 1 \Lra
{\bar\xi}_2 = e-1= 1.718$; then,
\begin{eqnarray}
\ln P_0[V] =
-N + 0.859141\,N^2 - 2.32178\,N^3 + 13.6475\,N^4 - \nonumber\\ 
  166.23\,N^5 + 4350.12\,N^6 - 257043.\,N^7 + 
  3.56058\,{10}^7\,N^8 - \nonumber\\ 1.18446\,{10}^{10}\,N^9 + 
  9.61505\,{10}^{12}\,N^{10} + {\Mfunction{O}(N)}^{11}.
\label{P0V-N}
\end{eqnarray}
When $N=0.2$, we have three decreasing terms, which yield $P_0[V] =
\exp(-0.184209) = 0.831762$, to be compared with the value computed directly
from Eq.\ (\ref{vpf-logn}), $P_0[V] = 0.838853$. Alternately, we can compare
it with the Poisson value $P_0[V] = \exp(-0.2) = 0.818731$.  If $N=0.01$, then
the series has more decreasing terms, namely, up to the seventh term. However,
the sum of those terms yields $-0.00991629$, that is to say, essentially the
same value as the first term, corresponding to a Poisson distribution.  Thus,
in the nonlinear regime, the sum of the decreasing terms of the series is a
very small correction to the first term $-N$ (the Poisson term).  It is clear
that the series expansion is then totally useless.

The non convergence of the cumulant expansion of the lognormal model is due to
the slow decay of its density probability distribution function in the high
density limit. In general, mass distributions that are singular on small
scales possess probability distribution functions with fat tails in the high
density limit.  If the singularities are not too strong, the probability
distribution function can have moments of any order, as the lognormal
distribution does. Then, the series expansions of the moment generating
function and of the void probability function are well defined, but they are
asymptotic rather than convergent.  Even when they exist, these cumulant
expansions are not useful for practical purposes in the nonlinear regime.

\subsection{Scaling in the nonlinear regime}
\label{nonlin-scaling}

In the strongly correlated regime ${\bar\xi}_2 \gg 1$, a systematic approach
is provided by the assumption of scale invariance (or self-similarity).  The
fractal regime is characterized by strong scaling correlations. In particular,
the mass fluctuations in a volume $V$ in a fractal follow a power law, namely,
${\bar\xi}_2 \sim V^{-\g/3} \gg 1$, where $\g$ is the scaling exponent of the
two-point correlation function, ${\xi}_2(r) \sim r^{-\g}$ (Gaite et al, 1999).
Then, the correlation dimension of the fractal is $3-\g$.  On the other hand,
it is natural to connect the VPF with the box-counting dimension: the former
gives the probability of a region of size $V$ (a box, say) being empty and the
latter gives the asymptotic number of non-empty boxes of size $V$ as $V \ra
0$. The box-counting dimension is equal to the correlation dimension in a
monofractal.

Let us take our sample to be defined in a region that we divide into a mesh of
cells with volume $V$ each.  Then, the ratios of empty or non-empty cells give
us estimations of $P_0[V]$ or $1-P_0[V]$, respectively.  We may call the
latter the non-void probability function.  Its behaviour as $V \ra 0$ is
related to the box-counting dimension of the distribution: the number of
non-empty cells of size $V$ is the power law $V^{-D_{\rm b}/3}$, where $D_{\rm
b}$ is the box-counting dimension and, therefore, the ratio of non-empty cells
is also a power law, namely, $V^{1-D_{\rm b}/3}$.  This exponent is always
non-negative and is zero only if $D_{\rm b} = 3$.  If the exponent is
positive, $P_0[V]$ tends to one. When $D_{\rm b} = 3$, $P_0[V]$ is not
constrained and the distribution can occupy any positive volume.

To analyse further the nonlinear behaviour of the VPF under the assumption of
scaling, we need a particular model.  Let us recall that the large deviation
formulation of multifractals (Harte, 2001) allows us to connect them with the
lognormal model, which can be regarded as a simple multifractal approximation
(Gaite, 2007).  We have the exact expression of the lognormal model VPF in
Eq.~(\ref{vpf-logn}).  Now, let us also recall the scaling behaviour of
density moments (Gaite, 2007):
$$
\mu_n(V) = \langle \r^n \rangle  \sim V^{(n-1)D_n/3 + D_0/3 - n},
$$ where $D_n$ are R\'enyi dimensions and, in particular, $D_0$ can be
identified with $D_{\rm b}$.  The exact expression of the lognormal model
moments is (Coles \& Jones, 1991)
$$
\langle \r^n \rangle = \exp\left(n\,\mu + \frac{n^2\s^2}{2}\right).
$$
Equating both expressions, we obtain
\begin{eqnarray}
\mu + \frac{\s^2}{2} \approx (D_0/3-1)\, \ln V\,, 
\label{mu-logn}
\\
n \frac{\s^2}{2} \approx (D_n-D_0) \,\ln V/3\,,
\label{s2-logn}
\end{eqnarray}
valid in the limit $V \ra 0 \Lra \mu,\s^2 \ra \infty$. If $\mu = -\s^2/2 \Lra
\langle \r \rangle = 1 \Ra D_0 = 3$, then a realization of the lognormal field
occupies the maximum volume and $P_0[V]$ vanishes. In general, the non-void
probability function $1-P_0[V] \sim \langle \r \rangle^{-1} \sim V^{1-D_0/3}$,
as explained above.

Let us consider the scaling lognormal model with $D_{\rm b} = 3$.  In the
strongly correlated regime, such that $\s \gg 1$ while $N$ stays finite,
$P_0[V]$ also approaches unity, according to Eq.~(\ref{vpf-logn}).  This limit
is a consequence of the strong clustering of particles, which leaves too large
voids, like when $D_{\rm b} < 3$.  The asymptotic behaviour $\s \ra \infty$
now yields
\begin{equation}
1-P_0[V] = \frac{\sqrt{2 N}}{\s}\, e^{-\s^2/8} + e^{-\s^2/8}\,
{\Mfunction{O}(\s^{-2})}.
\label{cvpf}
\end{equation}
Then, we deduce from Eqs.~(\ref{s2-logn}) and (\ref{cvpf}) that $1-P_0[V]$ is
a power law (except for a small logarithmic correction), namely,
\begin{eqnarray}
%1-P_0[V] \approx \frac{\sqrt{2 nV}}{\s}\, e^{-\s^2/8}
%\Ra \nonumber\\
\log(1-P_0[V]) \approx  \left(\frac{1}{2}+\frac{\g}{24}\right) \log V,
\label{log-log_cvpf}
\end{eqnarray}
where $\g = 3 - D_2$, again.

\begin{figure}
\centering{\includegraphics[width=7.5cm]{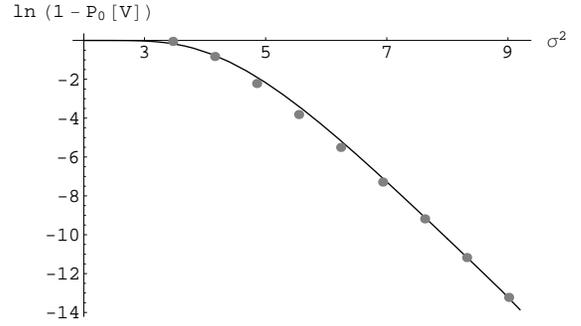}}
\caption{Log-log plot of the nonlinear lognormal non-void probability function
(note that $\s^2 \propto -\ln V$).  The paramenters used in this plot
correspond to the GIF2 cosmological simulation. The nine values are
estimations from counts in cells in the GIF2 simulation.
}
\label{lognormal_scaling}
\end{figure}

In Fig.~\ref{lognormal_scaling}, we have plotted $\ln(1-{P}_0[V])$, as given
by Eq.~(\ref{vpf-logn}), versus $\s^2 \propto -\ln V$.  We have used in the
calculation for that plot parameters corresponding to the GIF2 cosmological
simulation (Gaite, 2007): $n=400^3$, and $\s^2 = -(\g/3) \ln (V/r_0^3)$, where
$\g=1$ and $r_0=1/4$ is the scale of crossover to homogeneity in the
distributions of halos.  The scaling behaviour observed in the plot
corresponds to $\s^2 > 7$. Unfortunately, this behaviour is not given by Eq.\
(\ref{log-log_cvpf}) but it is the Poisson behaviour, because $N {\bar\xi}_2
<0 \Lra \s^2 > (3/\g -1)^{-1} \ln(nr_0^3) \simeq 7$.  Indeed, the Poisson VPF
for $N \ll 1$ is such that $\log(1-P_0[V]) = \ln(nV)$, which implies the
constant slope observed in the plot.  In Fig.~\ref{lognormal_scaling} also
appear nine values obtained by estimating ${P}_0[V]$ from the count-in-cell
analysis of the GIF2 simulation (Gaite, 2007).  We can appreciate that the
scaling lognormal model predicts correctly those values.

In the scaling lognormal model, all the R\'enyi dimensions $D_n$ are functions
of only two parameters, unlike in general multifractals, in which they are
independent.  However, Eqs.~(\ref{mu-logn}) and (\ref{s2-logn}) encompass
various types of fractal behaviour. For example, let us consider $\mu \ra
\infty$ while $\s$ stays finite. Then $D_n = D_0$ for every $n$, which
corresponds to monofractal behaviour. The non-void probability function
$1-P_0[V] \sim \langle \r \rangle^{-1} \sim V^{1-D_0/3}$ still vanishes, but
the fluctuations of the distribution in the non-void cells are bounded (they
can even be Gaussian).  If $\s \ra \infty$ as well, the distribution in the
non-void cells are unbounded, as in proper multifractals.  In any event, the
leading behaviour of the non-void probability function is given by $D_0$,
unless $D_0=3$.  We study the properties of general multifractal voids in
Sect.~\ref{MF}.

\section{Detecting voids: a spherical void finder}
\label{void-f}

The definition of voids in a point distribution is subtle and, to some extent,
subjective (except in a one-dimensional distribution, of course).  This
question has been amply discussed and various void-finders have been devised,
assuming different definitions of voids.  These definitions include
partially empty voids.  For example, the popular void-finder devised by El-Ad
\& Piran (1997) begins with a ``wall builder'', which separates ``field
galaxies'' from ``wall galaxies'' to delimit voids.  Then, it fits empty
spheres, but the final voids are unions of spheres with different radii and
have variable shape.  Nowadays, several void-finders are available,
mostly based on adaptable void shapes.  We have introduced a void-finder based
on discrete-geometry constructions that also defines voids of variable
shape and we have demonstrated that this void-finder is capable of
finding scaling in the void distribution (Gaite, 2005-B).

The separation of ``field galaxies'' that form a less clustered population is
a useful previous step, but its application is connected with the notions of
coexisting populations and galaxy biasing, which we study in detail in
Sects.~\ref{voids_unif-halos} and \ref{bias}. It is also connected with a
sophisticated type of voids, suggested by the scaling properties of
multifractals and introduced by Gaite (2007). Thys type of voids is studied in
Sects.~\ref{MF} and \ref{bias}.

In this section, we adopt the point of view of the preceding sections, based
on simple void shapes. For example, constant-shape voids are such that only
their size and orientation can change (apart from their position).  In
particular, if we use spheres, only their size can change.  Let us mention
that a widespread and more flexible alternative is the use of ellipsoids,
whose shape can change but is defined by few parameters. In fact, ellipsoids
are useful even as fits to more complicated shapes. However, their
eccentricity (departure from the spherical shape) needs to be bounded
above. Since the value of this bound is arbitrary, ellipsoids are less
suitable than spheres for a universal definition of voids.

As we consider that discrete-geometry constructions are the right starting
point, we first review the algorithm for variable-shape voids introduced in
Gaite (2005-B) and then we introduce a new and simple algorithm for spherical
voids, also based on discrete geometry constructions.

\subsection{Algorithm for variable-shape voids based on the Delaunay 
tessellation}

Given a set of isolated points, a natural geometric construction associated
with it is its Delaunay tessellation (Aurenhammer \& Klein, 2000; van de
Weygaert, 2002). This construction provides the {\em unique} set of largest
empty balls associated with the set of points, such that each ball is defined
as the circumscribed sphere to a Delaunay simplex (as a set of four
non-coplanar points).

Thus, the algorithm devised by Gaite (2005-B) was meant to be natural and
practical.  It begins with the Delaunay simplices and joins adjacent simplices
according to the overlap of their respective balls (Gaite, 2005-B). Therefore,
this void-finder provides a set of polyhedral voids that tessellate the entire
sample region; in other words, all the available space is assigned to
voids. In particular, some of the found voids can be very small: the total
range of sizes can span many orders of magnitude. This property is convenient
for testing the scaling of voids.  In fact, the void-finder demonstrates this
scaling (Gaite, 2005-B).

However, this void-finder only works properly in fractals with dimension $D$
larger than 2 (in three-dimensional ambient space).  If $D < 2$, the voids
tend to become degenerate, that is to say, tend to depart from round shapes,
and, when $D$ is sensibly smaller than two, it is likely that one void
percolates through the sample.  This percolation of voids can be avoided by
decreasing the overlap parameter, which controls the shape of voids (Gaite,
2005-B). Nevertheless, the value of the fractal dimension that the void-finder
yields is always larger than two, because the boundaries of the voids have
dimension two. This problem is better understood in terms of the notion of
cut-out sets (Gaite, 2006), which we study in Sect.\ \ref{cut-out}.  Of
course, other void-finders that define variable-shape and space-filling voids
can also yield a fractal dimension larger than two.

In general, the only way to prevent the ambiguity due to the shape of voids is
to prescribe voids of constant regular shape.  We have shown that void-finders
based on voids of constant shape demonstrate the scaling of voids in fractal
distributions; namely, the rank-ordering of the found voids fulfills
Zipf's power-law (Gaite \& Manrubia, 2002). Here, we propose an improved
void-finder of this type.

\subsection{New algorithm for finding spherical voids}
\label{new_v-f}

We have noticed that the scaling of voids is best realized when we impose that
voids touch the fractal, that is to say, when we reject the voids that do not
touch at least one point of the fractal (Gaite \& Manrubia, 2002).  Actually,
we should demand that the set of constant-shape voids has maximal contact with
the fractal.  In the case of spherical voids, we can demand that each sphere
is defined by four non-coplanar points on its boundary.  This condition
implies that each void is the sphere circumscribed to a simplex of the
Delaunay tessellation of the sample. Therefore, this tessellation is also the
primary element of a new and universal void-finder with no parameters.  We
further require that the balls are contained in the sample region and that
they do not overlap. Thus, the algorithm begins by finding the largest ball in
the sample region among those defined by the Delaunay tessellation, and
proceeds by searching for the next largest non-overlapping ball, until the
available balls are exhausted.

\begin{figure}
\centering{\includegraphics[width=7.5cm]{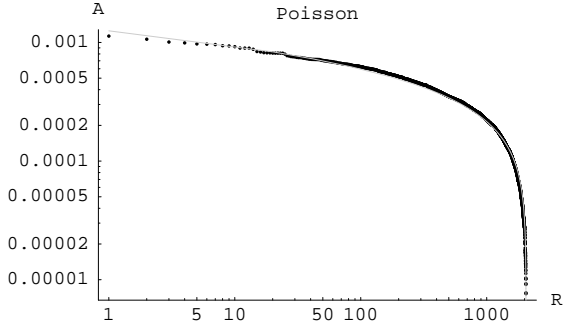}}
\caption{Rank-ordering of the circular voids in a random set of
$10\hspace{1pt}000$ points ($A$ is the void area and $R$ is the rank) compared
with the predicted law (gray line).}
\label{P-voids}
\end{figure}

Although we are interested in the scaling of voids, this void-finder is
applicable to any sample, in particular, to a sample of a uniform
distribution. In this case, we can test the laws studied in
Sect.~\ref{Poisson}. These laws refer to {\em all} the spherical voids in a
sample, but the no-overlap condition restricts the set of voids. However, the
sample of voids obtained under this condition is unbiased and, therefore, it
has the same distribution as the total set of voids.  To test it, we have
generated a random set of $10\hspace{1pt}000$ points in the unit square and
then run the void-finder. Its output (in rank order) is compared in
Fig.~\ref{P-voids} with the analytical prediction. This prediction results
from the two-dimensional version of the distribution of voids
$\mathbb{P}_0[V]$ given by Eq.~(\ref{cPP}) (replacing volume $V$ with area
$A$); namely, the rank is given by the cumulative probability $\mathbb{P}_>[A]
= (1 + nA)\, e^{-nA}$.  The agreement shown by Fig.~\ref{P-voids} is
remarkable.

\begin{figure}
\centering{\includegraphics[width=7.5cm]{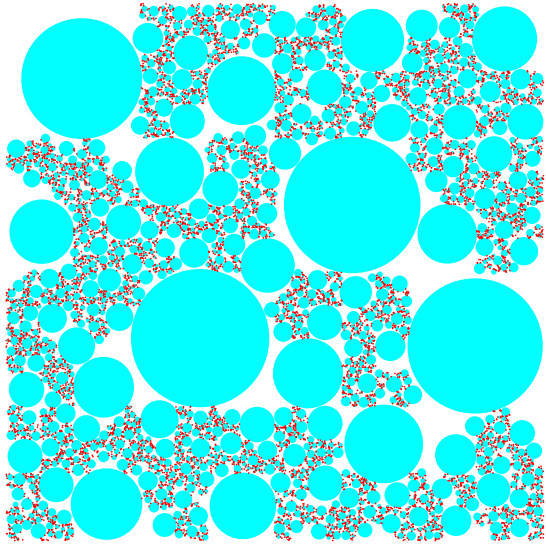}}\\[2mm]
\centering{\includegraphics[width=7.5cm]{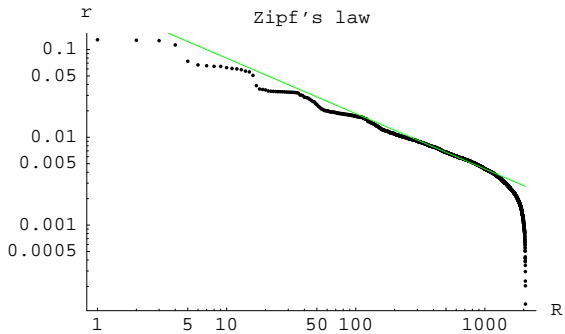}}
\caption{(Top figure) Random Cantor-like fractal sample with
$10\hspace{1pt}000$ points ($D=1.585$) and its corresponding voids found with
the new algorithm (described in the text). (Bottom figure) Log-log plot of the
rank-ordering of the void radii, compared with the straight line with slope
$1/D$.}
\label{blue}
\end{figure}

We have also applied this void-finder to several samples of random Cantor-like
fractals. We show in Fig.~\ref{blue} the results corresponding to a random
sample of $10\hspace{1pt}000$ points of a two-dimensional random Cantor-like
fractal with $D=1.585$. Note that the found circular voids (Fig.~\ref{blue},
top) do not cover the entire sample region (the unit square). In fact, they
cover only 70\% of it. Nonetheless, they convey well the notion of a hierarchy
of voids.

In addition, we have tested this void-finder on random Cantor-like fractals
that do not fulfill the condition $D > d-1$, where $d$ denotes the dimension
of the ambient space ($d=2,3$ in our tests).  While finders of space-filling
voids do not work or not find the right scaling when $D < d-1$, our new
void-finder works fine and obtains the approximate value of $D$. We explain
the theory of non space-filling voids in the next section.

Finally, let us remark that the algorithm is very simple and indeed runs very
fast.

\section{Scaling of fractal voids}
\label{scaling}

Scaling of voids is natural in a self-similar fractal: given that the fractal
is the union of a number of smaller similar copies of itself, every void also
has smaller similar copies of itself, such that there is an infinite hierarchy
of similar voids of decreasing size.  The simplest example of this similarity
of voids is the middle third Cantor set.  It is desirable to generalize this
argument to random self-similar fractals and to higher dimensions.

A self-similar distribution of voids fulfills the diameter-number relation
$N_>(\d) \simeq \d^{-D}$, namely, it is such that the cumulative number of
voids with diameter larger than a given value is a power law, with minus the
similarity dimension $D$ as exponent (Mandelbrot, 1983). Given that $N_>(\d)$
is the rank $R$ of the void with diameter $\d$, the rank-ordering of void
sizes is also a power law, that is, an instance of Zipf's law. A Zipf law is
usually demonstrated as a constant slope in a log-log plot.  The lengths of
the gaps in the middle third Cantor set follow Zipf's law; in particular, its
discrete scale invariance produces a staircase pattern in the log-log plot of
the rank-ordering of those lengths.  In random self-similar fractals, the
cumulative probability of sizes of voids also follows a power law.%
\footnote{Mandelbrot (1983) calls hyperbolic a power-law cumulative
distribution.  However, a more common name for it is Pareto distribution.}  In
three dimensions, scaling voids must fulfill the law $\mathbb{P}_>[V] \simeq
V^{-D/3}$.

The voids detected by various void-finders in random fractals indeed follow a
power law rank order (Gaite \& Manrubia, 2002; Gaite 2005-B).  In practice,
the Zipf law of voids constitutes a straightforward proof of fractality.
However, the absence of a formal definition of voids in dimensions larger than
one makes a rigorous study of the scaling of voids difficult. This was the
motivation for introducing the notion of cut-out sets (Gaite, 2006).

Cut-out sets are obtained by removing from an initial region an infinite
sequence of disjoint regions that exhausts the volume of the initial region
(Falconer, 1997).  These removed regions are the natural voids.  The final
structure is a sort of foam (under some conditions).  Therefore, cut-out sets
are relevant for ``cosmic foam'' models of large scale structure (Icke \& van
de Weygaert, 1987; van de Weygaert, 2002).

Fractals related to cut-out sets can be constructed with a modified procedure
that permits void merging, in such a manner that the voids form just one
connected region. For example, the fractal in Fig.~\ref{blue} is of this kind.
However, even more general fractals can have a meaningful sequence of voids,
such as the sequence found with our spherical void-finder. In particular, that
sequence can fulfill Zipf's law.  This motivates us to extend our previous
study of cut-out sets (Gaite, 2006) to more general fractals.

Connected with our generalization of cut-out sets is the notion of ``fat
fractal''.  Sets with fractal structure but non-zero volume were studied by
Mandelbrot (1983) and, later, they were called fat fractals (Grebogi et al,
1985).  We can distinguish two types of fat fractals: (i) non-scaling
fractals, namely, cut-out sets such that the sizes of their voids decrease
faster than a power law, and (ii) cut-out sets with scaling voids that do not
exhaust the initial volume.  The latter type is relevant for sequences of non
space-filling voids, which we study in Sect.~\ref{non-filling}.

In this section, we restrict ourselves to totally empty voids in continuous
distributions. However, a finite fractal sample has, in addition, Poissonian
fluctuations that can give rise to Poissonian voids, as studied in
Sects.~\ref{Poisson} and \ref{correl}.  We postpone the study of Poissonian
voids to Sect.~\ref{MF}.

\subsection{Cut-out sets and scaling of voids}
\label{cut-out}

Fractal cut-out sets are obtained by removing an infinite sequence of disjoint
connected open regions from an initial compact and convex region, in such a
way that the sum of the removed regions tends to the volume of the initial
region.%
\footnote{Let us recall some basic geometrical notions that are necessary
here.  Given a region in Euclidean space, its boundary is formed by points
such that any ball centered on them intersects both the region and its
complement.  A closed set contains its boundary. If a closed set is bounded,
it is also compact.  An open set does not contain any boundary point. The
complement of an open set is closed and vice versa. The union of open sets is
open.  A disconnected set can be divided into two parts such that each one is
disjoint with the boundary of the other.  A convex set contains every segment
with ends in the set.}  
We can consider the voids in order of decreasing size and that every void is
cut out from the remainder of the previous cuts.  A well known one-dimensional
example is the middle third Cantor set.  In a sense, every closed fractal is a
cut-out set, because its complement is open and, therefore, it is the union of
a sequence of disjoint connected open regions.%
\footnote{This statement is a classic theorem of topology, proved by, e.g., 
Franz (1965).} 
However, there may only be a finite number of them. We regard as proper 
cut-out sets the closed sets with vanishing volume and an 
infinite sequence of cutouts (voids).

In one dimension, voids are necessarily open intervals. Thus, every closed
one-dimensional fractal can be constructed like the Cantor set.  In higher
dimensions, a connected open region can have a very complicated shape; for
example, it can have any number of ``matter islands'' and it can have a very
rough boundary (a familiar image of a connected open region is provided by the
shape of a cloud). Therefore, it is convenient to restrict ourselves to
regular shapes and, in particular, to convex voids (Gaite, 2006). Under this
condition and the condition that the voids do not degenerate to lower
dimensional objects (along the sequence), the Zipf law of voids holds.

Let us mention a famous class of deterministic cut-out fractals with convex
voids in two dimensions: the Sierpinski carpets. The triadic Sierpinski carpet
is constructed as a sort of two-dimensional generalization of the middle third
Cantor set: from an initial square, the middle (open) sub-square of side one
third is cut out, and the iteration proceeds with the remaining eight
sub-squares. It has $D = 1.89$.  Variants of this construction produce other
cut-out fractals with square voids (Mandelbrot, 1983). It is straightforward
to generalize these constructions to three dimensions. All these fractals have
discrete scale invariance and the log-log plots of the rank-orderings of their
void sizes produce typical staircase patterns.

In contrast, fractals without discrete scale invariance can be generated by
either deterministic or random algorithms, and are such that the size of their
voids tends to decrease in a continuous fashion. For example, in the random
fractal in Fig.~\ref{blue}, the (relatively smooth) steps on small ranks
vanish on larger ranks.  In general, we can express the scaling of voids as a
particular power-law form of the rank order of diameters: $\d(R) \asymp
R^{-1/D_{\rm b}}$, in terms of the relation $\asymp$, which means that the
quotient between the related quantities is bounded above and below.  This
number-diameter relation is equivalent to a common form of Zipf's law: the
log-log plot of the rank ordering stays between two parallel lines with a
slope given by the exponent ($-1/D_{\rm b}$, in the present case).  Of course,
if the sequence of voids is non-degenerate ($V \asymp \d^3$), we can replace
$\d$ with $V^{1/3}$ in the rank-ordering, and
\begin{equation}
V(R) \asymp R^{-3/D_{\rm b}}.
\label{V-R}
\end{equation}

A cut-out fractal is formed by the union of the boundaries of its voids.%
\footnote{To be mathematically rigorous, a cut-out set is the boundary of the
union of its voids rather than the union of the boundaries of its voids. The
former is usually larger than the latter and is actually its closure, namely,
the smallest closed set that contains it. This subtle difference is irrelevant
regarding its box-counting dimension but is relevant regarding its
Hausdorff-Besicovitch dimension.}
Naturally, a cut-out set has fractal dimension larger than two (in
three-dimensional space).%
\footnote{A rigorous proof of this fact involves the notion of topological
  dimension (Mandelbrot, 1983).}
Cut-out sets with convex and non-degenerate voids formalise the geometry of
fractal foams.  For example, the voids can be convex polyhedra, like in the
Voronoi foam model of Icke \& van de Weygaert (1987).  Besides, the mass may
not be homogeneously distributed on the boundaries of voids.  For example, the
Voronoi foam model of Icke \& van de Weygaert (1987) is based on the expansion
of initial under-dense regions, which become depleted while the walls between
them concentrate their mass.  However, the evolution of the resulting foam
continues with the motion of the matter in the walls towards their
intersections to form filaments, and, then the motion along the filaments to
form nodes.

\begin{figure}
\centering{\includegraphics[width=7.5cm]{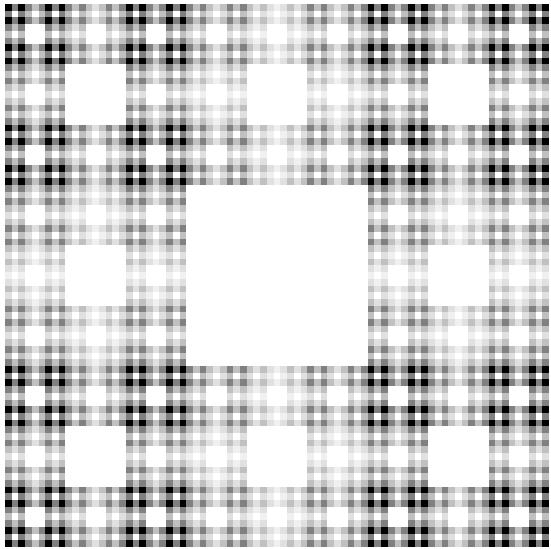}}
\caption{Two-dimensional inhomogeneous fractal foam: the Cantor-Sierpinski
carpert. It is a multifractal cut-out set: beside the empty voids
corresponding to the Sierpinski carpet, we can perceive very low density
regions, in contrast with the mass concentrations near the two-dimensional
Cantor dust.}
\label{C-S}
\end{figure}

To illustrate inhomogeneous cut-out sets, we can use a toy model inspired in
the Sierpinski carpet, which we call the {\em Cantor-Sierpinski carpet}. We
construct it with a slightly modified Sierpinski algorithm.  The first step of
the fractal generator consists in cutting out the middle sub-square of side
one third from an initial square. We can describe this operation as a uniform
displacement of mass from the middle sub-square to the surrounding sub-squares
(as in a model of cosmic foam). Then, following the above-mentioned idea of a
latter mass displacement along walls, we further concentrate part of the mass
in the four sub-squares at the corners. Thus, the generator consists in the
following way of dividing the total mass into the nine sub-squares: the
central one receives nothing, the four sub-squares at the corners each receive
a proportion $p_1$ of the total, and the remaining four sub-squares each
receive a proportion $p_2 < p_1$, with $4 (p_1 + p_2) = 1$. The resulting
inhomogeneous cut-out fractal is supported on the Sierpinski carpet but has
maximal mass concentrations on the two-dimensional Cantor dust (the Cartesian
product of two Cantor sets). The case $p_1 = 1/6,\,p_2 = 1/12$ is shown in
Fig.~\ref{C-S}.  In fact, the Cantor-Sierpinski carpet is a multifractal,
which we revisit in Sect.~\ref{MF}.

Now, let us consider a different modification in the construction of a cut-out
set.  It is easy to see that the boundaries of different voids can intersect
one another. Then, it may be possible to remove a piece of common boundary so
that the joined voids form a larger void, namely, a connected open region
(which is not necessarily convex).  With this merging prescription, the shapes
of the voids become more complex and, in fact, all the voids may form a finite
number of connected regions, or even only one region.  Moreover, the modified
(improper) cut-out set may not contain any piece of surface. A two-dimensional
example of complete merging of voids is the fractal in Fig.~\ref{blue}.  An
ideal void-finder should be able to undo the merging, but there is no general
procedure to recover a unique cut-out set structure and, hence, a unique set
of voids.

On the other hand, if the dimension of a (closed) fractal is smaller than two,
then it cannot be constructed as a (proper) cut-out set: its complement (in
the region where the fractal is defined) is connected.%
\footnote{We assume that the region where the fractal is defined is a given
simple region, preferably convex. By default, we assume a cube. If no natural
region is available, we can take the {\em convex hull} of the fractal, namely,
the smallest convex set that contains it.}  Then, in the absence of a natural
void structure, it does not make sense to try to find ``real'' fractal voids
by using a space-filling void-finder, and it may be more sensible to look for
equal-shaped voids, as we have argued in Sect.\ \ref{void-f}.  Therefore, we
are going to explore how to generalize the relation between the scaling of
voids and the fractal dimension to voids that touch a fractal but do not
totally fill its complementary region (like the circular voids in a square
shown in Fig.~\ref{blue}).

\subsection{Scaling and dimension provided by non space-filling voids}
\label{non-filling}

Let us consider a sequence of voids that touch a fractal but do not totally
fill the fractal's complement.  If every point of the fractal touches a void,
the fractal is included in the boundary of the voids.  Mandelbrot (1983)
studied self-similar unions of boundaries, especially, in two dimensions,
under the name of sigma-loops (sigma-loop = sum of loops). As self-similar
objects, they are constructed by means of a generator, constituted by a number
of parts; each part is substituted by a copy of the whole (scaled down by some
factor). To such construction one can associate the similarity dimension
$D_{\rm s}$, namely, the quotient of the logarithm of the number of parts by
the logarithm of the inverse of the similarity ratio.  Mandelbrot (1983)
measures every loop by its linear scale $\d$ (its diameter) and shows that the
number of loops fulfills the diameter-number relation $N_>(\d) \asymp
\d^{-D_{\rm s}}$. The generalization to three dimensions (in terms of
sigma-surfaces) is straightforward.

If a fractal is not exactly self-similar or if it does not coincide exactly
with the boundary of its voids, Mandelbrot's treatment is not
applicable. However, the method of Falconer (1997), extended by Gaite (2006),
can be applied with some modifications.  Let us recall that the method of
Falconer (1997) is based on the equality of the box-counting dimension $D_{\rm
b}$ and the Minkowski-Bouligand dimension. The latter dimension expresses the
power-law behaviour, as $r \ra 0$, of the volume $V(r)$ of the
$r$-neighbourhood of the fractal, which is the union of the balls of radius
$r$ with centers in the fractal.  The rank-ordering of voids provides bounds
to $V(r)$, hence allowing one to connect the decrease rate of the volumes of
voids with the decrease rate of $V(r)$.

Following the proof for cut-out sets, we can try to carry out the proof for
general convex voids, instead of only equal-shaped voids (or balls).  Let us
assume the rank order of the diameters of the voids to be $\d(R) \asymp
R^{-a}$, with exponent $a$.  We look for bounds to $V(r)$ in terms of the
voids.  The lower bound to $V(r)$ is independent of whether or not the voids
are space-filling: the $r$-neighbourhood of the fractal is certain to contain
every void with diameter smaller than $r$, because the voids touch the
fractal.  Then, the total volume of the voids fully included in that
$r$-neighbourhood provides a lower bound to $V(r)$ that implies $D_{\rm b}
\geq 1/a$.

To obtain an upper bound to $D_{\rm b}$, we need an upper bound to the volume
$V(r)$.  However, the method of Gaite (2006) for cut-out sets needs
modifications, because a good part of the $r$-neighbourhood of the fractal is
now not included in the voids but in its complement (a fat fractal).  Instead,
we can appeal to the more general (but more complex) results of Tricot (1986,
1989). Tricot proves that the open sets packing the complement of the fractal
can be quite general. This generality implies that the scaling of voids is not
necessarily connected with the box-counting dimension $D_{\rm b}$, like in
cut-out sets, but it can be connected with a different type of dimension, the
so-called {\em exterior capacity} dimension, which can be more appropriately
called here the interior dimension of the voids.

Finally, let us notice that our algorithm does not guarantee that every point
of the fractal belongs to the boundary of a void: each initial ball is defined
by four points on its boundary, but many balls are removed to satisfy the
non-overlap condition.  The points of the fractal not included in the boundary
of the voids, namely, the points that do not touch voids, can also make a
contribution to the fractal dimension, if their relative weight is
non-negligible, increasing the difference $D_{\rm b} - 1/a \geq 0$.
Nonetheless, the equality $D_{\rm b} = 1/a$ seems to hold for random
self-similar fractals.  Indeed, our tests on simulations show that the
algorithm yields the right dimension, even when the fractal dimension is
smaller than two.

\section{Voids in a multifractal}
\label{MF}

Mandelbrot (1983) was concerned with the size and aspect of fractal voids in
the distribution of galaxies. In fact, he favoured small voids in this
distribution.  Thus, he introduced the concept of fractal lacunarity and
proposed that the galaxy distribution is a fractal with low lacunarity.
Mandelbrot (1983) actually showed that fractals with the same dimension can
look very different, according to their lacunarity.  Finally, he presented a
brief study of the Besicovitch fractal, which has been later described as a
self-similar binomial multifractal.  Mandelbrot (1983) called it instead a
``non-lacunar'' fractal, alluding to the fact that it has no open voids.  The
modern literature about multifractals is not particularly concerned with their
voids, but the structure of multifractal voids indeed has interest in
cosmology.

In a multifractal, the mass surrounding a point $x$ grows as a power law with
an exponent that varies with the point, $\a(x)$; namely, $m[B({x},r)] \simeq
r^{\a(x)}$, where $m[B(x,r)]$ is the mass in the ball of radius $r$ centered
on ${x}$.  This definition makes sense for points such that 
$m[B(x,r)]$ is nonzero for all $r > 0$. These points form the support of the
distribution, which is necessarily a closed set.%
\footnote{Indeed, the support of a mass distribution is precisely defined as
the smallest closed set that contains all the mass.}  A multifractal possesses
a spectrum of dimensions, such as the set of local dimensions $\a(x)$ or the
set of R\'enyi dimensions $D(q),\,-\infty < q < \infty$ (Harte, 2001).  The
box-counting dimension $D_{\rm b}$ of the multifractal support can be
identified with $D(0)$ .

A {\em monofractal} is defined as the particular case in which the
local dimension $\a$ is constant throughout its support.  
In other words, a monofractal is a uniform mass distribution with fractal 
support.%
\footnote{To be precise, the uniformity {\em only} affects the exponent. 
The pre-factor in the power law can vary, giving rise to different 
monofractal distributions with the same support and exponent, but with 
different (bounded) inhomogeneities in the support.}
Of course, the dimension of the support coincides with the
local dimension $\a$.

The notion of a void must be more complicated in multifractals than in
monofractals, because of the spectrum of dimensions.  To study in detail the
nature of voids in multifractal distributions, we need to reconsider the
definition of fractal voids. We can actually distinguish two types of voids.

\subsection{The two types of voids in a multifractal}

In Sect.~\ref{scaling}, we have defined a void as a connected open region
(which it is useful to consider convex). However, a multifractal can have a
different type of voids. To introduce it, let us consider the example of the
one-dimensional adhesion model. In this model, the mass concentrates in {\em
shocks} and their locations form a countable but {\em dense} set (She, Aurell
\& Frisch, 1992; Vergassola et al, 1994); namely, in any interval, however
small, there are shock points. Since the set of shock points is countable, it
has null length, so the the total length of the interval actually belongs to
its complementary set.  Therefore, this complementary set contains no mass and
has non-vanishing length, but it does not contain any open void, because the
shock points are dense.

Thus, it is too restrictive to require that voids be open. Indeed, in the
adhesion model, every finite sample of the mass distribution displays voids
(She, Aurell \& Frisch, 1992; Vergassola et al, 1994).  The mass distribution
in the adhesion model is actually of multifractal type. A general self-similar
multifractal mass distribution also has an infinite number of mass
concentrations, with local dimension $\a(x) < 3$ and diverging
density. Moreover, these singular mass concentrations are dense in the support
of the mass distribution; namely, any open region (or any ball) contains mass
concentrations.  Among the regular points, with a well-defined density, the
points that we naturally assign to voids are the ones with local dimension
$\a(x) > 3$ and vanishing density (Gaite, 2007). Thus , the set of points with
vanishing density has non-vanishing volume (indeed, it usually holds the total
volume), but it may not contain any open set.

Therefore, we are going to distinguish two types of voids in general
continuous mass distributions.  The first type of voids consists in open
regions. There can be one, many or an infinite sequence of void open connected
regions.  Their complementary region is closed and constitutes the
distribution support. If there are no voids of that type, in other words, if
the distribution support occupies the total volume in which the distribution
is defined, we must consider a more general type of voids: the set of points
with vanishing density. Of course, the density is also zero in open voids, but
we demand that voids of the second type belong to the support of the mass
distribution.  Both types of voids are present in some self-similar
multifractals, for example, in the Cantor-Sierpinski carpet (Fig.~\ref{C-S}).

In general, multifractal distributions may or may not have voids of the first
type but they certainly have voids of the second type, which are the proper
multifractal voids.  However, the large density fluctuations inherent to
multifractals imply that the distinction between both types of voids gets
blurred in finite samples, in which the voids are Poissonian.  This remark is
important, of course, for the application to real data. Then, the only way to
determine the type of voids is by increasing the sample density $n$, so that
voids of the second type get filled.

When we consider finite samples, a coarse-grained version of the mass
distribution is adequate to study the geometry of voids. Let us see that the
distinction between the two types of voids disappears in coarse-grained mass
distributions.

\subsection{Connection with excursion sets}

Self-similar multifractal mass distributions have a dense set of singular mass
concentrations where the density diverges. To remove these singularities, it
is convenient to coarse-grain the distribution, for example, using a window
function or low-pass filter of wave-numbers. Then the mass density is well
defined everywhere.  The lognormal model, already employed in
Sect.~\ref{correl}, is a suitable coarse-grained approximation to self-similar
multifractal mass distributions (Gaite, 2007).  A density field allows us to
define voids as the regions where the density is below some given
threshold. This definition has been introduced by Sheth \& van de Weygaert
(2004) and by Shandarin, Sheth \& Sahni (2004).

On the other hand, a coarse-grained continuous distribution can be obtained
from a finite sample and, if the coarse-graining length is well chosen, it is
a good description of it. Regarding voids, the appropriate coarse-graining
length is such that, in average, there is one particle in a volume of that
length ($N=1$). Thus, under-dense regions are real voids (Gaite, 2007).

A density field resulting from coarse-graining is continuous (in general).
This implies that excursion sets are open sets, if they are defined as the
points where the density is strictly smaller than a given value (say the
average density).  Therefore, their geometry is similar to the geometry of
multifractal voids of the first type.  As we have already mentioned, every
open set is formed by a sequence of connected open regions. Every connected
open region constitutes an individual void. This geometry is simpler than the
geometry of multifractal voids of the second type. However, let us emphasise
that a connected open region can be very complex.

\begin{figure}
\centering{\includegraphics[width=7.5cm]{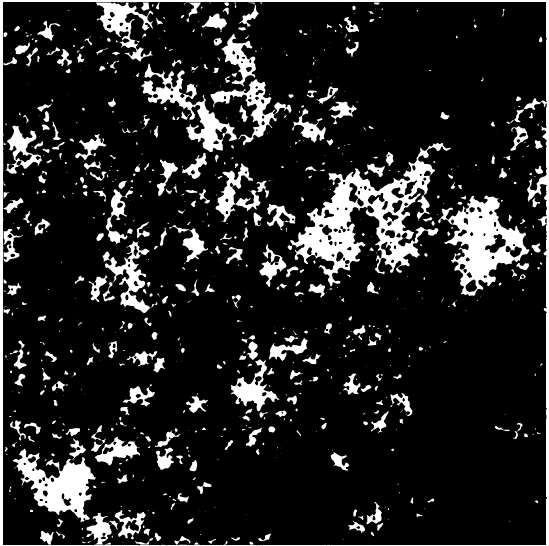}}
\caption{Voids (in black) defined by the average-density excursion set 
of a lognormal model.}
\label{2dFBM}
\end{figure}

For illustration, we have computed excursion sets of a realization in a square
of a two-dimensional lognormal field with $\langle\r\rangle = 1$ and $\s =
1.65$.  We have plotted in Fig.~\ref{2dFBM} the excursion set corresponding to
the average density (one).  The black region is the excursion set, which can
be decomposed into a set of connected regions that constitute individual
voids.  A good part of the total volume belongs to the largest void, which
percolates through the square.  There are smaller voids as islets inside the
matter clusters (``voids in clouds''). Unlike in a Gaussian field, there is no
symmetry between clusters and voids.  The matter clusters contain most of the
mass but have small total volume.  Indeed, the matter clusters occupy only
20\% of the volume but hold 80\% of the total mass.  Since the matter clusters
occupy a small but non-vanishing volume, we can regard them as a fat
fractal.

The largest connected void in Fig.~\ref{2dFBM} illustrates the complicated
geometry that a connected open region can have (a geometry that can be even
more complicated in three dimensions). It seems natural to divide a void like
that into smaller but simpler regions. A natural way to do it is by choosing
the smaller regions convex, like we did in cut-out sets.  Then, we understand
that the total connected region is constructed by merging the smaller convex
regions.  However, a careful look at Fig.~\ref{2dFBM} will convince us that
the necessary number of convex components of the largest void is huge.

From Fig.~\ref{2dFBM}, we can imagine the geometry of voids of the second
type.  When the coarse-graining length decreases, more and more matter halos
pop up in the voids and more and more voids pop up in the matter clusters, as
the mass distribution becomes more singular.  In an $N$-body cosmological
simulation, Gottl\"ober et al.\ (2003) re-simulated voids with higher
resolution and indeed observed the formation of small halos in them in a
self-similar pattern.  In the limit of vanishing coarse-graining length, halos
(mass concentrations) are fully mixed with (are dense in) some parts of the
voids, which form voids of the second type.  The open voids that may remain
constitute voids of the first type. Moreover, voids occupy an increasing
fraction of the total volume that tends to one, and contain a decreasing
fraction of the total mass that tends to zero.

\subsection{Scaling of voids in a multifractal}
\label{MF_voids}

We have shown in Sect.~\ref{scaling} that fractal voids of the first type
follow a diameter-number relation (under some conditions) and its exponent is
given by the box-counting dimension $D_{\rm b}$ of the fractal.  A
multifractal possesses a spectrum of dimensions; in particular, $D_{\rm b}$
can be identified with the R\'enyi dimension $D(0)$.  In
Sect.~\ref{nonlin-scaling}, we have related $D_{\rm b}$ to the VPF.  If
$D_{\rm b} = D(0) < 3$, then the support of the distribution is fractal, and
the volumes of the corresponding first type voids scale according to
Eq.~(\ref{V-R}).  However, when $D(0) = 3$, the voids cannot satisfy
Eq.~(\ref{V-R}): it would tell us that the total volume of the voids $\sum_R
V(R)$ diverges.

Actually, the case $D(0)=3$ is particularly important in our context, because
the mass distributions obtained in $N$-body simulations of CDM models are
consistent with $D(0)=3$ (Gaite, 2007). Furthermore, the supports of those
distributions seem to be their entire regions of definition, such that
there are no voids of the first type.  However, this conclusion is far from
being certain.  To confirm it, it is necessary to analyse simulations with
larger ratios of the homogeneity scale to the discretization scale.

Regarding voids of the second type (non open), they have a very complicated
geometry, as we have explained above.  Indeed, the geometry of such voids is
hard to describe.  There can be an uncountable number of connected components
or only one.  In any event, it is very difficult to establish the sizes of
separate voids and they may not be rank ordered. In particular, if there is an
uncountable number of connected voids, it is possible that every separate void
has zero volume, in spite that their total volume is positive.  This happens
in the one-dimensional adhesion model.

However, the radical differences between second type voids and open voids
disappear when we consider a finite fractal sample.  A finite sample naturally
concentrates in regions with local dimension $\a(x) < 3$ (halos), whereas
voids are depleted (Gaite, 2007). Therefore, one can perceive more or less
regular void shapes in a multifractal sample, even in the absence of voids of
the first type. In fact, a finite multifractal sample can be described in
terms of a coarse-grained mass distribution and its voids can be described in
terms of excursion sets.  However, we have seen that excursion sets are still
complex.  Therefore, it is convenient to define a sequence of convex voids by
means of a void-finder; in particular, we can use the sequence of spherical
voids found with the algorithm in Sect.~\ref{void-f}.  We have studied in
Sect.~\ref{nonlin-scaling} scaling features related to the VPF.  We study the
scaling of multifractal voids below.

\subsection{Poissonian voids in a multifractal}
\label{P-MFvoids}

For simplicity, we consider here only multifractals supported in their entire
domains of definition, that is to say, such that they only have voids of the
second type. A discrete sampling of a continuous distribution gives rise to
Poissonian voids, as we studied in Sect.~\ref{correl}.  Since the perturbative
methods in that section are not applicable to multifractals, we study here the
distribution of the voids in simulated random self-similar multifractals. In
particular, we study a random self-similar multifractal with $D(0) = 3$ and no
open voids, employing the void-finder described in Sect.~\ref{new_v-f}.

We analyse two-dimensional random multinomial multifractals (Harte, 2001).  We
define a particular multifractal in the unit square, with great precision;
namely, we define about $2.8\,10^{14}$ ``pixels'' (allowing us to specify
coordinates with seven decimal digits). This multifractal has support in the
whole unit square and, therefore, it has no voids of the first type.  We have
generated a sample of this multifractal with $10\hspace{1pt}000$ points (the
typical order of magnitude of galaxy VLS's). The application of our
void-finder to this sample yields some (Poisson) voids with relatively large
size (see Fig.~\ref{blue-666}). For example, the largest void has a radius
equal to 0.0484 and an area equal to 0.00736 (in box-size units). According to
the results of Sect.~\ref{Poisson}, the expected number of voids of that size
in a sample of the uniform distribution with $10\hspace{1pt}000$ points, such
that $N = 73.6$, would be $10\hspace{1pt}000\,73.6^2\, \exp(-73.6) \simeq
6\,10^{-25}$. Naturally, this small number shows how inhomogeneous this
multifractal is. Regarding the distribution of voids, the log-log plot
rank-ordering of the sequence of void radii does not fulfill Zipf's law (see
Fig.~\ref{blue-666}).

We have also tested samples with different numbers of points.  Smaller
samples, namely, with less than $10\hspace{1pt}000$ points, have larger voids,
but the total number of voids decreases.  Of course, the rank-orderings of
these reduced sequences of voids do also not follow a Zipf law. Larger samples
have more voids, with smaller size (which can be very small).  Therefore,
larger samples are more influenced by the error due to the finite precision of
the multifractal measure.  We have not observed any scaling range.  In fact,
the aspect of the log-log plots for all these rank-orderings (e.g.,
Fig.~\ref{blue-666}, bottom plot) is not unlike the aspect of the plot that
corresponds to the Poisson field (Fig.~\ref{P-voids}).

Regarding the high number density of dark matter particles, totally empty
spherical voids in their distribution must be very small. Therefore, it is
more practical to consider spherical voids in the distributions of mass
concentrations (halos), rather than in the raw mass distribution, following
the ideas proposed by Gaite (2005-A, 2007). Voids in the distribution of halos
can be related to voids in the distribution of galaxies, as we explain later.

\subsection{Voids in a sample of uniform halos}
\label{voids_unif-halos}

According to the conclusions of Gaite (2007), some scaling properties of a
self-similar multifractal may not be realized even in large samples. For
example, the scaling of the two-point correlation function is not realized in
$N$-body CDM simulations of current size (many millions of
particles). Consequently, it is necessary to carefully select the most
adequate scaling quantities. In particular, a well-motivated selection
consists in the choice of quantities corresponding to a uniform halo
population, namely, corresponding to a given local dimension $\a < 3$.  If
halos are realized by coarse-grained lumps of a given size, a uniform halo
population consists of halos with similar mass. We expect that the voids in
each population have better scaling properties, like do the correlation
functions of each population (Gaite, 2005-A, 2007).  We use the preceding
multifractal to test this hypothesis.

\begin{figure}
\centering{\includegraphics[width=7.5cm]{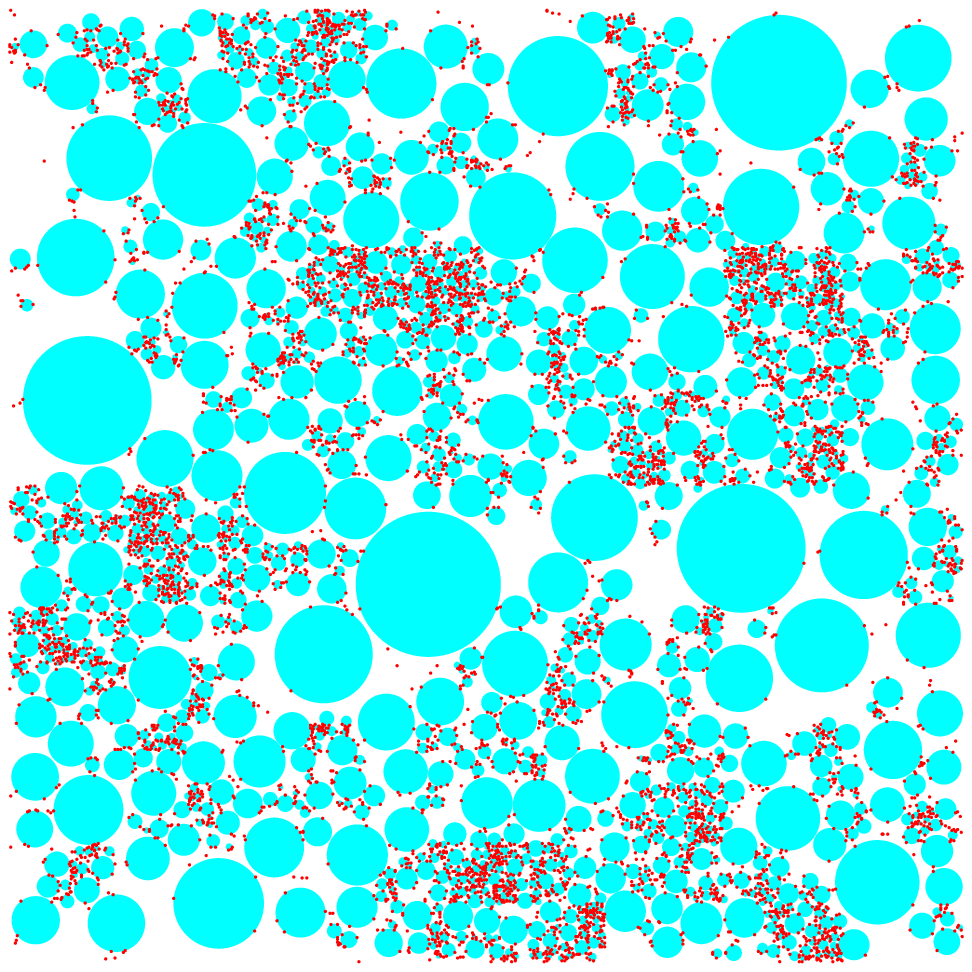}}\\[2mm]
\centering{\includegraphics[width=7.5cm]{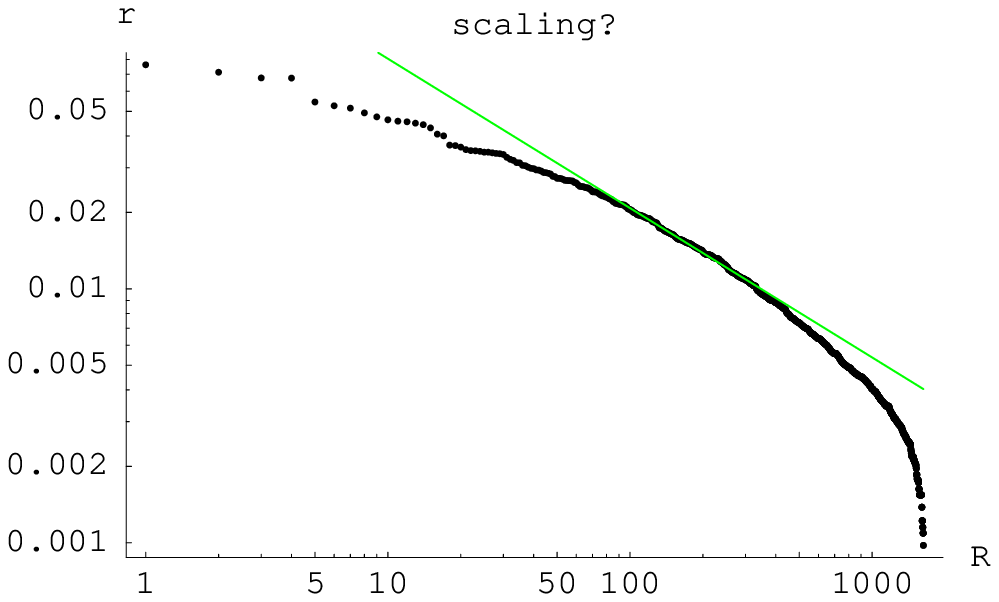}}
\caption{Set of $7\hspace{1pt}548$ multifractal halos with masses between 450
and 550 particles, and its corresponding voids (top figure).  Log-log plot of
the rank-ordering of the void radii (bottom plot), in comparison with the 
straight line required by the Zipf law (corresponding 
to the dimension $D = 1.7$).}
\label{blue-667}
\end{figure}

To do the test, we need a sample with many more than $10\hspace{1pt}000$
points to coarse-grain them into a sensible number of halos. Therefore, we
have generated a sample with $67\hspace{1pt}108\hspace{1pt}864$ points. We
have coarse-grained it in a $1\hspace{1pt}024 \times 1\hspace{1pt}024$
mesh. We have selected the halos with a number of particles between 450 and
550, corresponding to $\a \simeq 1.7$; the resulting number of halos is
$7\hspace{1pt}548$.%
\footnote{The value $\a \simeq 1.7$ is slightly smaller than $\a_1 \simeq
1.73$, the local dimension of the mass concentrate of this distribution. This
local dimension is, in general, such that 
$\a_1 = f(\a_1)$, that is to say, such that it coincides with 
the Hausdorff-Besicovitch dimension of the concentrate.}  Then we
have proceeded with these halos like with ordinary particles, namely, we have
applied our void-finder and studied the void distribution. The results are
shown in Fig.~\ref{blue-667}.

\begin{figure}
\centering{\includegraphics[width=7.5cm]{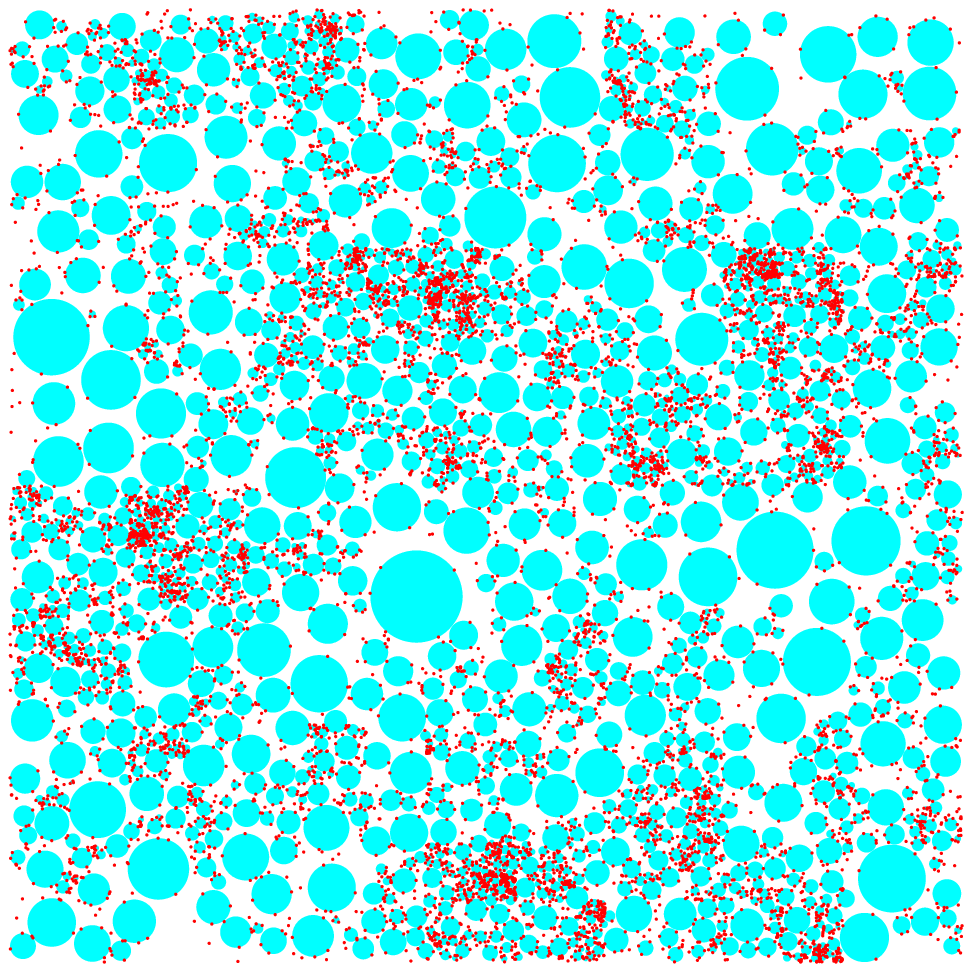}}\\[2mm]
\centering{\includegraphics[width=7.5cm]{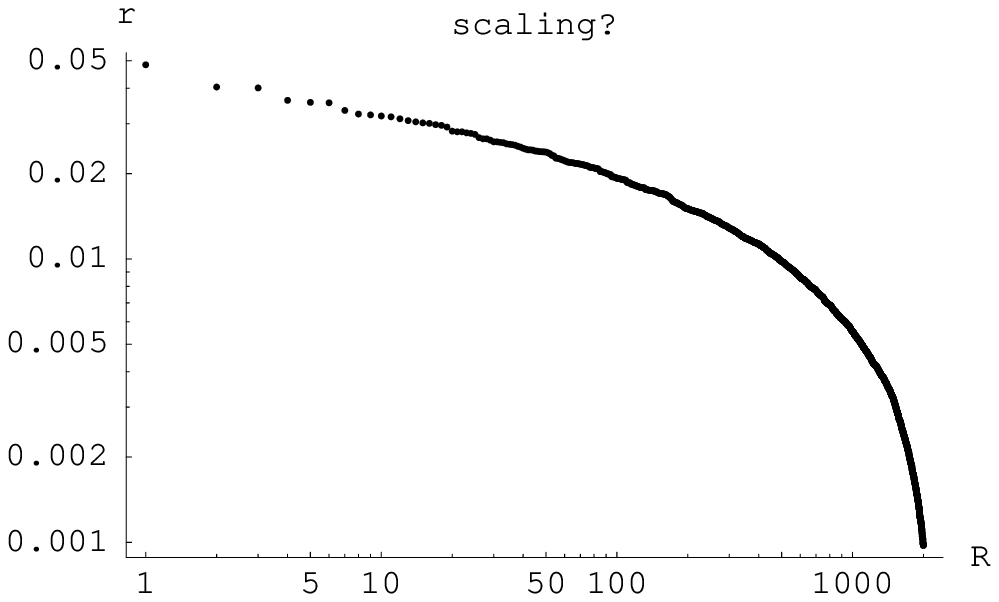}}
\caption{Random multifractal sample with $10,000$ points and its
corresponding voids (top figure) found with the new algorithm. 
Log-log plot of the rank-ordering of the void radii (bottom plot).}
\label{blue-666}
\end{figure}

The voids in this set of halos are larger than the voids in the raw particle
distribution (see Fig.~\ref{blue-666}). The reason for it is double: the set
of halos is more clustered than the raw particle distribution, and, also,
there are fewer halos now ($7\hspace{1pt}548$) than particles in the above
sample ($10\hspace{1pt}000$).  Now, the largest void has a radius equal to
0.0763. Regarding all the voids, the largest voids are now larger but they are
located in about the same places.  The sequence of voids now spans a larger
size range, and its distribution is a little different from that of the voids
of the basic particle population: a tentative fitting of a Zipf law yields a
better result (Fig.~\ref{blue-667}). However, it is questionable that our
selection of a uniform halo population approaches scaling behaviour. This
result is to be contrasted with the clear scaling of correlation functions
that is achieved by the selection of uniform halo populations (Gaite,
2005-A, 2007).

Let us notice the relation of the above selection procedure to the ``wall
builder'' phase in the void-finder of El-Ad \& Piran (1997). Indeed, their
procedure also separates (galaxy) populations according to their clustering,
although in a less discriminatory way: it only separates lowly-clustered
``field galaxies'' from highly-clustered ``wall galaxies.''  Actually, their
criterion for the separation of galaxies should coincide with a sort of
discrete adaptation of a simplified version of our criterion for halos: to
separate two populations according to their values of $\a$, we must just
prescribe a threshold, such that the values below it are lowly-clustered while
the values above it are highly-clustered.  Since the local dimension $\a(x)$
measures the concentration of mass around $x$, the mass in the ball of radius
$r$ centered on ${x}$ is $m[B({x},r)] \sim r^{\a(x)}.$ A discrete version of
$m[B(x,r)]$ is given by the number of points inside $B(x,r)$.  A threshold for
$\a$ sets a threshold for this number, as El-Ad \& Piran do.

\section{Voids in galaxy samples and galaxy bias}
\label{bias}

Let us recall the observational evidence of scaling of sizes of galaxy voids,
in particular, the study of voids in the 2dF survey by Tikhonov (2006)
mentioned in Sect.~\ref{Tikho} regarding the significance of large voids.
Tikhonov performs the rank-ordering of voids in the mentioned VLS and indeed
concludes that there is a scaling range.  This range is about a decade,
namely, an order of magnitude in the rank (from rank 60 to rank 600,
approximately).  Tikhonov (2006) uses his own void-finder, which first fits
the largest empty spheres and then applies a merging criterion that allows the
voids to become non-spherical (in a similar way to El-Ad \& Piran, 1997).

\begin{figure}
\centering{\includegraphics[width=7.5cm]{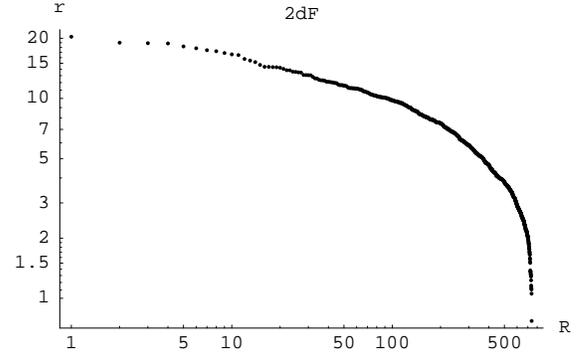}}
\caption{Rank-ordering of the voids in Tikhonov's 2dFGRS sample with
$7\hspace{1pt}000$ galaxies ($r$ is the void radius in Mpc.\ $h^{-1}$).}
\label{2dF-voids}
\end{figure}

The void-finder defined in Sect.~\ref{void-f} is simpler, for it only finds
empty spheres, without merging them.  We have applied our void-finder to
Tikhonov's VL sample.\footnote{Actually, we have removed a few galaxies from
one boundary to make it straight, thus making the geometrical shape of the
sample rectangular (in the angular coordinates). Furthermore, we have shifted
slightly the position of the straightened edge in order to have a round number
of galaxies in the sample, namely, $7\hspace{1pt}000$ (out of the initial
$7\hspace{1pt}219$).}  The resulting rank order is plotted in
Fig.~\ref{2dF-voids}. The range of radii is similar to the range found by
Tikhonov, although somewhat larger at the lower end (which is not
significant).  However, no scaling range can be discerned in that plot.

We must also consider that previous analysis of galaxy catalogues have not
revealed scaling of sizes of voids (Gaite \& Manrubia, 2002). Indeed, it has
not been clear if we should really expect this scaling.  Our argument for
scaling (Gaite \& Manrubia, 2002) actually assumed a monofractal distribution
of galaxies.  Therefore, the voids in it should be of the first type, in our
present classification.  However, regarding that the distribution of galaxies
is better described as a multifractal, the scaling of voids can be more
complex, as we have discussed above.  To draw conclusions for galaxy voids
from our study of multifractal voids, we need to consider in detail the
relation between the dark matter and galaxy distributions.

\subsection{Multifractal model of galaxy bias}

One might assume that galaxies just trace the dark matter distribution,
namely, that they constitute a fair sample of this distribution, considered
continuous (``galaxies trace mass'').  Any more sophisticated prescription
amounts to galaxy biasing. Galaxy biasing is natural, since the principles
that rule the distribution of galaxies are complex (indeed, the formation of
galaxies is not well understood yet).  There are various models of galaxy
biasing.  Our model combines the ideas of the ``peak theory'' of Gaussian
fields (Kaiser, 1984; Bardeen et al, 1986) with our multifractal halo model
(Gaite, 2005-A, 2007).
 
In a Gaussian field, the density is well defined everywhere. In contrast, in a
singular distribution, in particular, in a multifractal, the density is not
defined (it is, actually, infinite) in a large set of points.  These points
are mass concentrations, characterized by their local dimensions $\a(x)$.
Therefore, we must substitute the density threshold employed to define peaks
of Gaussian fields by a local-dimension threshold.  This substitution becomes
an actual equivalence if we coarse grain the singular distribution: then, the
coarse-grained density measures the local-dimension. However, the
coarse-grained density field is not Gaussian while the coarse-graining length
is small (in the nonlinear regime).

Of course, the natural threshold is $\a(x) < 3$, which is just the definition
of halos as mass concentrations proposed by Gaite (2005-A, 2007).  The local
dimension $\a=3$ corresponds to a coarse-grained density similar to the
average density (in the support of the distribution).  Therefore, the simplest
model of galaxy biasing consists in that every halo hosts a galaxy with mass
(or luminosity) proportional to the mass of the halo.  Thus, the distribution
of galaxies of given luminosity has a well defined fractal dimension.  An
important consequence of this model is that the total distribution of galaxies
is also multifractal.  This consequence can be tested in galaxy correlation
functions.  In fact, Zehavi et al (2005) and Tikhonov (2006) have found that
the slope $\g$ of the log-log plot of the two-point correlation decreases with
luminosity in a way that agrees with a multifractal distribution of galaxies.

The value of the threshold is relevant for the definition of ``wall
galaxies'', as we comment in Sect.~\ref{voids_unif-halos} regarding
halos.  Assuming that every halo hosts a galaxy, we can set a lower $\a$
threshold (higher density threshold) for ``wall galaxies''. A natural choice
is the local dimension $\a_1$ of the mass concentrate, such that $\a_1 =
f(\a_1)$, which is the fractal (Hausdorff-Besicovitch) dimension of the
concentrate. This choice ensures that the mass of all the ``field galaxies'',
corresponding to $\a_1 < \a <3$, is almost negligible.

In any event, if every halo hosts a galaxy with luminosity proportional to the
mass of the halo, the voids in a population of galaxies of given luminosity
coincide with the voids in the parent halo population. Therefore, we can
extend the conclusions of the study of voids in samples of uniform halos
performed in Sect.~\ref{voids_unif-halos} to galaxy voids; namely, the voids
(of the second type) in them span a large range of sizes and their
rank-ordering is vaguely similar to a Zipf's law.

For the sake of completeness, let us mention a more elaborate model of
galaxies in dark matter halos that assumes that a halo can host more than one
galaxy (see, e.g., Peacock and Smith, 2000).  However, if the distribution of
galaxies in a given halo follows the distribution of the dark matter in it,
this model should be equivalent to the simpler model with only one galaxy per
halo, provided that the dark matter distribution is multifractal.  The reason
for this equivalence is that the size of halos in the multifractal halo model
is given by the chosen coarse-graining length, which can run within some
limits without altering anything, due to the scale invariance (Gaite,
2007). Therefore, an increase in the size of halos must indeed correspond to
placing more galaxies per halo, without altering their distribution.

Of course, the present observational limitations do not allow us to fully test
these models or to determine the nature of galaxy voids. For example, we
cannot tell how much matter the present voids contain or exactly in what form.
In particular, small dark matter halos in voids are undetectable.  Galaxy
VLS's are naturally biased towards the more luminous populations.  Thus, their
voids do contain galaxies. In general, the observed voids seem to contain
dwarf or low surface brightness galaxies.  Peebles (2001) discusses the nature
of void objects.

Finally, let us note that the various rank-orderings of voids in galaxy VLS's
available agree qualitatively with a multifractal model of galaxy bias;
namely, the corresponding log-log plots look like those in Fig.~\ref{blue-666}
or Fig.~\ref{blue-667}.  Nevertheless, it is remarkable that some analyses
seem to really demonstrate scaling, unlike our plots.  This could hint at a
more complicated model of galaxy biasing or at a dark matter distribution with
totally empty scaling voids.

\section{Discussion and Conclusions}
\label{discuss}

The traditional Poissonian analysis of voids, based on the perturbative void
probability function, is only valid when the Poisson fluctuations dominate
over the correlations ($N\bar\xi_2 < 1$). Thus, if there is more than one
object per volume of the size of the homogeneity scale ($N> 1$ when
$\bar\xi_2 =1$), then the Poissonian analysis is only valid in the nonlinear
regime. In particular, the Gaussian approximation can only be used, if at all,
for very sparsely distributed objects, like Abell clusters or very luminous
galaxies.

In the nonlinear regime, when the perturbative Poissonian analysis is valid,
it yields no information. To obtain information in the nonlinear regime, we
assume scale invariance of the correlation functions. Then, essentially two
different situations are possible, according to the behaviour of the void
probability function in the continuous distribution, namely, in the limit of
infinite sampling density ($n \ra \infty$).  When the box-counting dimension
of the distribution is smaller than three, the void probability function
approaches unity as the cell volume $V \ra 0$.  This happens, for example, in
a monofractal.  Conversely, if the void probability function of the continuous
distribution does not approach unity as $V \ra 0$, then the box-counting
dimension of the distribution is three.  Moreover, if the void probability
function vanishes, a random cell is surely non-empty and normal voids are
absent in the continuous distribution.  In other words, voids are present only
while the sampling density $n$ is finite (Poissonian voids).  This Poissonian
void probability function approaches unity as $V \ra 0$ and is related to a
different R\'enyi dimension.  The various behaviours of the void probability
function are well illustrated by the scaling lognormal model.

Relying on previous work and for the sake of analytical simplicity, we have
used empty spherical voids.  Thus, we can carry out a complete analysis of
voids in the Poisson distribution and a partial analysis in correlated
distributions.  We have designed a new and simple finder of non-overlapping
spherical voids.  We have tested it on Poisson distributions and ordinary
fractal distributions, obtaining the expected rank-orderings of voids, namely,
the analytical Poisson law and the Zipf law, for a random point distribution
and a random Cantor fractal, respectively.  The random fractal illustrates the
aspect of spherical voids in a continuous mass distribution: they do not fill
the void space but constitute a good approximation to a a partition of it.

Focusing on continuous distributions, the scaling of voids is best studied by
introducing the notion of cut-out sets. This notion is very general and every
monofractal is, in a sense, a cut-out set.  Mandelbrot's diameter-number
relation can be translated into a power-law rank-ordering of void sizes
(Zipf's law).  Cut-out sets with non-degenerate convex voids formalise the
geometry of fractal foams.  Non-convex voids can be formed from convex voids
by a process of merging. However, given a non-convex void, it cannot be
partitioned into a unique set of convex components.

In contrast, spherical voids (balls) do not tessellate a part of space, unless
we fill the interstices with smaller balls (forming an {\em Apollonian
packing} of balls). However, to really represent the structure of the voids in
a given distribution, we demand (in our void-finder) that the balls touch the
points of the distribution, while some space remains unfilled by the sequence
of voids. Thus, the complement of the voids constitutes a fat fractal. The
scaling of voids then is less straightforward and can involve a different
dimension (the interior dimension of the voids instead of the box-dimension of
the mass distribution).  However, spherical voids are usually a good
approximation to a partition of the empty space into non-degenerate convex
regions.  In particular, voids scale with the right exponent in our
simulations of random self-similar fractals.

Multifractal mass distributions have a spectrum of dimensions and contain
singular mass concentrations of variable strength. Therefore, multifractal
voids are more complex, but they can actually be classified in two types,
which correspond to the two main behaviours of the void probability function
in the limit $V \ra 0$.  Voids of the first type are like monofractal voids,
only present when the void probability function tends to one.  In contrast,
voids of the second type are formed by mass depletions, typical of
multifractals, which are present even when the void probability function
vanishes. This type of voids appears, for example, in the adhesion model and
involves complex geometrical notions.  Our classification of voids provides an
interpretation of Mandelbrot's lacunarity concept: either type of voids
characterizes lacunar or non-lacunar fractals, respectively.  

We have illustrated the difference between both types of voids with a
deterministic multifractal foam: the Cantor-Sierpinski carpet.  The ``cosmic
web'' can be modelled as a random multifractal foam.  However, the results of
simulations of cold dark matter dynamics are consistent with the presence of
the second type of voids only.

Multifractal geometry is complex and not intuitive. Notwithstanding, a
coarse-grained mass distribution and, therefore, its coarse-grained voids are
sufficient to describe finite multifractal samples.  Both types of
multifractal voids mix under coarse graining, becoming excursion sets. In
nonlinear (non-Gaussian) fields, the excursion set that defines voids occupies
most of the volume but contains little mass. In particular, in the lognormal
model, that excursion set is dominated by a percolating void.

After coarse graining, connected multifractal voids still have complicated
structures.  It is convenient to partition them into simpler regions, for
example, convex regions, like we do in cut-out sets.  In particular, spherical
voids (balls) are adequate for finite multifractal samples. We have studied
the distribution of spherical voids in samples of simulated multifractals
supported in their whole regions of definition (with voids of the second type
only).  The log-log plots of the rank-orderings of void sizes are similar to
those of Poisson distributions, but the sizes of small voids do not
decrease as sharply as in the latter.  The voids in distributions of uniform
halos are more relevant for the dark matter distribution. The sizes of these
voids decrease quite smoothly, without actually scaling.

Regarding voids in the galaxy distribution, we employ a multifractal model of
galaxy bias, which we propose as a multifractal version of the ``peak theory''
of Gaussian fields. It has a free parameter, the halo mass threshold for
``wall galaxies''.  It is natural to choose a high threshold. Assuming that
this threshold is higher than the threshold for galaxy formation in halos,
there are galaxies in the voids (``field galaxies'' or ``void
galaxies''). Otherwise, voids do not contain galaxies. Nevertheless, voids
defined as excursion sets or spherical voids defined by galaxy samples contain
mini-halos with a substantial amount of dark matter.

\begin{acknowledgements}
I thank Rien van de Weygaert and the other organizers of the Royal Academy
Colloquium ``Cosmic Voids'' in Amsterdam, Dec.~2006, for the invitation to
participate in it, where the idea for this work arose.  I also thank Anton
Tikhonov for sending me his 2dF VLS data and for correspondence, Fernando
Barbero for calculating the asymptotic expansion in Eq.\ (\ref{cvpf}), and
Claude Tricot for sending me some references.
\end{acknowledgements}

\end{document}